\documentclass[12pt,eqno,epsf]{article}

\usepackage{amsmath,amssymb,graphicx,epsf}

\numberwithin{equation}{section}

\def\ignore#1{{}}

\newcounter{sxn}

\newcounter{axn}

\date{}

\newdimen\mybaselineskip
\renewcommand{\thefootnote}{\arabic{footnote}}

\newcommand{\beeq}{\begin{equation}}
\newcommand{\eneq}{\end{equation}}
\newcommand{\beqn}{\begin{eqnarray}}
\newcommand{\eeqn}{\end{eqnarray}}

\newcommand{\alp}{\alpha}
\newcommand{\bt}{\beta}
\newcommand{\gm}{\gamma}
\newcommand{\Gm}{\Gamma}
\newcommand{\dlt}{\delta}
\newcommand{\Dlt}{\Delta}
\newcommand{\ep}{\epsilon}
\newcommand{\tht}{\theta}

\newcommand{\Tht}{\Theta}
\newcommand{\vth}{\vartheta}

\newcommand{\Lmd}{\Lambda}
\newcommand{\sgm}{\sigma}

\newcommand{\omg}{\omega}

\newcommand{\be}{\begin{equation}}
\newcommand{\ee}{\end{equation}}
\newcommand{\bea}{\begin{eqnarray}}
\newcommand{\eea}{\end{eqnarray}}
\newcommand{\eql}{\!\!\!&=\!\!\!&}

\newcommand{\sma}{\!\!\!&\simeq\!\!\!&}
\newcommand{\defa}{\!\!\!&\equiv\!\!\!&}

\newcommand{\toa}{\!\!\!&\to\!\!\!&}

\newcommand{\tl}[1]{\tilde{#1}}
\newcommand{\bdm}[1]{{\mbox{\boldmath $#1$}}}

\newcommand{\der}{\partial}

\newcommand{\ie}{{i.e.}}

\newcommand{\brkt}[1]{\left( #1 \right)}
\newcommand{\brc}[1]{\left\{ #1 \right\}}
\newcommand{\sbk}[1]{\left[ #1 \right]}
\newcommand{\abs}[1]{\left| #1 \right|}

\newcommand{\fl}[1]{\lfloor #1 \rfloor}

\renewcommand{\Re}{{\rm Re}\,}
\renewcommand{\Im}{{\rm Im}\,}

\newcommand{\cF}{{\cal F}}
\newcommand{\cG}{{\cal G}}

\newcommand{\cL}{{\cal L}}

\newcommand{\cO}{{\cal O}}

\newcommand{\cR}{{\cal R}}

\begin{document}
\thispagestyle{empty}

\baselineskip=12pt


\begin{flushright}
KEK-TH-2376 \\
WU-HEP-21-04
\end{flushright}

\baselineskip=25pt plus 1pt minus 1pt

\vskip 1.5cm

\begin{center}
{\LARGE\bf UV sensitivity of Casimir energy}

\vspace{1.0cm}
\baselineskip=20pt plus 1pt minus 1pt

\normalsize

{\large\bf Yu Asai}${}^1\!${\def\thefootnote{\fnsymbol{footnote}}
\footnote[1]{E-mail address: u-asai.physics@ruri.waseda.jp}} 
{\large\bf and Yutaka Sakamura}${}^{2,3}\!${\def\thefootnote{\fnsymbol{footnote}}
\footnote[2]{E-mail address: sakamura@post.kek.jp}}

\vskip 1.0em

${}^1${\small\it Department of Physics, Waseda University, \\ 
3-4-1 Ookubo, Shinjuku-ku, Tokyo 169-8555, Japan}

\vskip 1.0em

${}^2${\small\it KEK Theory Center, Institute of Particle and Nuclear Studies, 
KEK, \\ 1-1 Oho, Tsukuba, Ibaraki 305-0801, Japan} \\ \vspace{1mm}
${}^3${\small\it Department of Particles and Nuclear Physics, \\
SOKENDAI (The Graduate University for Advanced Studies), \\
1-1 Oho, Tsukuba, Ibaraki 305-0801, Japan}

\end{center}

\vskip 1.0cm
\baselineskip=20pt plus 1pt minus 1pt

\begin{abstract}
We quantitatively estimate the effect of the UV physics on the Casimir energy 
in a five-dimensional (5D) model on $S^1/Z_2$. 
If the cutoff scale of the 5D theory is not far from the compactification scale, 
the UV physics may affect the low energy result. 
We work in the cutoff regularization scheme by introducing 
two independent cutoff scales for the spatial momentum in the non-compact space 
and for the Kaluza-Klein masses. 
The effects of the UV physics are incorporated as a damping effect of the contributions 
to the vacuum energy around the cutoff scales. 
We numerically calculate the Casimir energy and evaluate the deviation from 
the result obtained in the zeta-function regularization, which does not include information on the UV physics. 
We find that the result well agrees with the latter  
for the Gaussian-type damping, while it can deviate for the kink-type one. 
\end{abstract}

\newpage

\section{Introduction}
The Casimir effect is a well-known macroscopic quantum effect~\cite{Casimir:1948dh}. 
It has been observed by various experiments, 
and the observed values well agree with the theoretical 
predictions~\cite{Lamoreaux:1996wh}-\cite{Bimonte:2021sib}. 
The Casimir effect can also play an important role in the context of 
the extra dimensions. 
It generates the scalar potential for the volume modulus of the compact extra space, 
and can stabilize the modulus to a finite value~\cite{Garriga:2000jb}-\cite{Nojiri:2000bz}. 

The Casimir energy is defined as the energy difference between the vacuum state in the presence of the conducting plates 
and that in the absence of the plates.  
Since the vacuum energies in quantum theories generically diverge, we have to regularize them before taking the difference. 
There are various ways to calculate the Casimir energy with different regularization schemes,  and the same results are obtained 
in the renormalizable theories 
by those methods~\cite{Boyer:1968uf}-\cite{Fichet:2021xfn}. 
However, it is also known that there can be a mismatch 
between the results obtained using different regularizations in some cases~\cite{Beneventano:1995fh}-\cite{Matsui:2018tan}. 
Such discrepancies come from the (regularized) divergent part of the vacuum energies. 
Among various regularizations, the cutoff regularization is presumably the most physically intuitive way to regularize the divergent quantities. 
Ref.~\cite{Visser:2016ddm} clarified the conditions that the Casimir energy becomes the cutoff independent, 
and is finite. 
It also pointed out that the zeta-function regularization 
discard the cutoff dependence that may have some physical information on the UV physics. 
For the four-dimensional (4D) electrodynamics, the fact that the experimental values are in agreement with 
the theoretical predictions indicates that such cutoff dependences are negligible. 
However, this does not ensure that they can always be negligible in any theories.


The superstring theory is a promising candidate of the final theory that contains the quantum gravity, 
and  predicts the existence of the extra space dimensions. 
Such extra space is often supposed to be compactified on some manifolds or orbifolds 
in order to explain the fact that the observed space dimension is three. 
So if we detect any signal that suggests the extra dimensions, it strongly supports the superstring theory. 
The existence of the compact extra space affects the time evolution of the universe 
because it modifies the Einstein equation. 
Therefore, the search for the deviation from the standard cosmological evolution is useful to test the superstring theory. 
The current cosmological obervations indicate that the spacetime evolves the Freedman equation that oiginates from 
the 4D Einstein equation. 
Thus the size of the compact space should be stabilized at a small finite value, 
so that its effects on the cosmological evolution of the universe are suppressed at late times. 
In our previous work~\cite{Abe:2020jnp}, we investigated the time evolution of the domain-wall configuration in the $S^1$ extra dimension, 
which can be regarded as the 3-brane we live. 
In that setup, the extra dimension continues to expand due to the repulsive force between the kinks. 
Hence we need some moduli-stabilization mechanism in order to obtain the (approximate) 4D FLRW universe at late times. 
As we mentioned, the Casimir effect can be used for such stabilization. 
In fact, if the Casimir force is attractive, it can balance with the repulsive force between the kinks. 

Since the extra-dimensional models are non-renormalizable, 
it should be regarded as effective theories of more fundamental theories. 
In other words, they have UV cutoff scales. 
In this paper, we discuss a five-dimensional (5D) scalar theory compactified on $S^1/Z_2$ as a simple example. 
The theory behaves as a 5D theory between the cutoff scale~$\Lmd$ 
and the compactification scale~$1/R$, where $R$ is the radius of $S^1$. 
In particular, when the cutoff~$\Lmd$ is not far from $1/R$, effects of the UV physics 
might give non-negligible contributions to the result. 
Such effects will appear as a cutoff-dependence of the Casimir energy. 

The purpose of this work is to quantitatively estimate the UV-cutoff dependence of the Casimir energy, 
and to clarify the situation in which it is non-negligible.\footnote{
The cutoff-dependence of the Casimir energy has been also discussed in Ref.~\cite{Mahajan:2006mw} 
with a different motivation. 
They discussed a possibility that the cosmological constant originates from the Casimir energy 
in the 4D context, and focused on a case that the spacing between the plates, which corresponds 
to the size of the extra dimension~$\pi R$ in our case, is shorter than the cutoff length~$1/\Lmd$. 
Another different point from their work is that we take into account effects of the momentum cutoff, 
which is in principle independent of the cutoff for the Kaluza-Klein masses. 
} 
For our purpose, we work in the cutoff regularization scheme, 
do not take the limit~$\Lmd\to\infty$, and numerically evaluate the deviation from the result 
obtained by the conventional methods, which corresponds to the limit of $\Lmd\to\infty$. 


The paper is organized as follows. 
In the next section, we provide a brief review of the conventional calculations 
for the Casimir energy. 
In Sec.~\ref{cutoff_reg}, we introduce the cutoff scales for the momentum 
and the Kaluza-Klein masses, 
and derive the expression of the Casimir energy. 
In Sec.~\ref{calc_Cas_eng}, we numerically calculate the Casimir energy, 
and evaluate the deviation from the conventional result. 
Sec.~\ref{summary} is devoted to the summary and discussions. 
In the appendices, we collect some formulae used in our calculations.

\section{Conventional calculations for Casimir energy}
In this section, we provide a brief review of the conventional way to calculate 
the Casimir energy~\cite{Lamoreaux:1996wh}-\cite{Brevik:2000vt}. 
To simplify the discussion, we consider a real scalar theory 
in the flat $D$-dimensional spacetime,\footnote{
We leave the spacetime dimension~$D$ unspecified in this and the next sections. 
In numerical calculations performed in Sec.~\ref{calc_Cas_eng}, we focus on the case of $D=5$. 
} 
and one of the spatial dimensions is compactified on $S^1/Z_2$.  
\bea
 \cL \eql -\frac{1}{2}\der^\mu\Phi\der_\mu\Phi-\frac{1}{2}M^2\Phi^2, 
\eea
where $\mu=0,1,\cdots,D-1$, $M$ is the bulk mass parameter. 
The coordinate of the compact dimension is denoted as $y$. 
The fundamental region of $S^1/Z_2$ is chosen as $0\leq y\leq \pi R$, 
where $R$ is the radius of $S^1$. 
The real scalar field~$\Phi$ is assumed to be $Z_2$ odd. 
Thus, it satisfies the Dirichlet boundary conditions at the boundaries of $S^1/Z_2$, 
in addition to the periodic boundary condition. 
As a result of these boundary conditions, the Kaluza-Klein (KK) masses are determined by
\be
 \sin\brkt{\pi R\sqrt{m_n^2-M^2}} = 0, 
\ee
whose solutions are 
\be
 m_n = \sqrt{M^2+\frac{n^2}{R^2}}. \;\;\;\;\; \brkt{n=1,2,\cdots}
\ee
Then, the vacuum energy density in the $(D-1)$-dimensional effective theory 
is expressed as
\be
 E_{\rm vac} = \sum_{n=1}^\infty \int\frac{d^dk}{2(2\pi)^d}\;\omg_n(k), 
 \label{V_eff:1}
\ee
where $d\equiv D-2$ is the dimension of the non-compact space, and 
\be
 \omg_n(k) \equiv \sqrt{k^2+m_n^2} \;\;\;\;\; \brkt{k^2\equiv |\vec{k}|^2}
\ee
is the energy of a KK mode with the $d$-dimensional momentum~$\vec{k}$ and the mass~$m_n$. 

To perform the $\vec{k}$-integral, we work in the dimensional regularization, and obtain
\bea
 E_{\rm vac} \eql 
 -\frac{\Gm(-\frac{d+1}{2})}{2(4\pi)^{(d+1)/2}}\sum_{n=1}^\infty m_n^{d+1}, 
 \label{V_eff:DR}
\eea
where $\Gm(\alp)$ is the Euler gamma function. 
The infinite sum over the KK modes is evaluated 
by means of the zeta function regularization technique~\cite{Goldberger:2000dv,Leseduarte:1996ah,Leseduarte:1996xr}. 
Using the formula~(\ref{final_expr:Sp:2}), the above expression becomes
\bea
 E_{\rm vac} 
 \eql -\frac{\Gm(-\frac{d+1}{2})}{2(4\pi)^{(d+1)/2}R^{d+1}}\left\{-\frac{\bar{M}^{d+1}}{2}
 +\frac{\sqrt{\pi}\Gm(-\frac{d+2}{2})}{2\Gm(-\frac{d+1}{2})}\bar{M}^{d+2} \right.\nonumber\\
 &&\hspace{35mm}\left.
 +\frac{2\bar{M}^{\frac{d+2}{2}}}{\pi^{\frac{d+1}{2}}\Gm(-\frac{d+1}{2})}
 \sum_{n=1}^\infty n^{-\frac{d+2}{2}}K_{\frac{d+2}{2}}\brkt{2\pi n\bar{M}}\right\} \nonumber\\
 \eql \frac{\Gm(-\frac{d+1}{2})M^{d+1}}{4(4\pi)^{(d+1)/2}}
 -\frac{\Gm(-\frac{d+2}{2})}{8(4\pi)^{d/2}}RM^{d+2} \nonumber\\
 &&-\frac{M^{\frac{d+2}{2}}}{(2\pi)^{d+1}R^{d/2}}
 \sum_{n=1}^\infty n^{-\frac{d+2}{2}}K_{\frac{d+2}{2}}\brkt{2\pi n MR}, 
 \label{conventional_E_vac}
\eea
where $\bar{M}\equiv MR$. 
The first term is irrelevant to the Casimir force because it is independent of $R$. 
The second term gives a constant Casimir force, which is also present even in the case of $R\to\infty$. 
Since the physically relevant Casimir force is 
the difference between the above $E_{\rm vac}$ and that of the non-compact case (\ie, $R\to\infty$), 
the contribution from the second term in (\ref{conventional_E_vac}) is cancelled. 
Thus, the Casimir energy~$E_{\rm cas}$ is given by 
\be
 E_{\rm cas} = -\frac{M^{\frac{d+2}{2}}}{(2\pi)^{d+1}R^{d/2}}
 \sum_{n=1}^\infty n^{-\frac{d+2}{2}}K_{\frac{d+2}{2}}\brkt{2\pi nMR}. 
 \label{conventional_E_cas}
\ee
In the massless case ($M=0$), (\ref{conventional_E_cas}) reduces to 
\bea
 E_{\rm cas}^{(M=0)} \eql -\frac{1}{2^{d+1}\pi^{\frac{3}{2}d+2}R^{d+1}}
 \sum_{n=1}^\infty \frac{1}{n^{d+2}}\lim_{M\to 0}\brkt{\pi n MR}^{\frac{d+2}{2}}
 K_{\frac{d+2}{2}}\brkt{2\pi nMR} \nonumber\\
 \eql -\frac{1}{2^{d+1}\pi^{\frac{3}{2}d+2}R^{d+1}}\sum_{n=1}^\infty \frac{1}{n^{d+2}}
 \frac{\Gm(\frac{d+2}{2})}{2} \nonumber\\
 \eql -\frac{\Gm(\frac{d+2}{2})\zeta(d+2)}{2^{d+2}\pi^{\frac{3}{2}d+2}R^{d+1}}, 
\eea
where $\zeta(s)$ is the Riemann zeta function. 
We have used the formula: 
\be
 \lim_{z\to 0}z^{\alp}K_\alp(2z) = \frac{\Gm(\alp)}{2}. \;\;\;\;\; \brkt{\alp>0} 
\ee
Namely, we have 
\be
 E_{\rm cas}^{(M=0)} = \begin{cases} 
 \displaystyle -\frac{\zeta(3)}{16\pi^3R^2} & \brkt{d=1} \\
 \displaystyle \rule{0mm}{24pt} -\frac{1}{1440\pi R^3} & \brkt{d=2} \\
 \displaystyle \rule{0mm}{24pt} -\frac{3\zeta(5)}{128\pi^6R^4} & \brkt{d=3} \\
 \displaystyle \rule{0mm}{24pt} -\frac{1}{30240\pi^2R^5} & \brkt{d=4} \end{cases}. 
 \label{value:conventional_E_cas}
\ee

The Casimir energy~(\ref{conventional_E_cas}) can be also expressed 
in the following integral form by using the formula~(\ref{int_form:Sp}). 
\bea
 E_{\rm cas} \eql -\frac{\Gm(-\frac{d+1}{2})}{2(4\pi)^{(d+1)/2}R^{d+1}}\cdot
 \frac{d+1}{\pi}\sin\frac{(d+1)\pi}{2}\int_0^\infty dw\;w^d
 \ln\brkt{1-e^{-2\pi\sqrt{w^2+\bar{M}^2}}} \nonumber\\
 \eql \frac{1}{(4\pi)^{(d+1)/2}\Gm(\frac{d+1}{2})R^{d+1}}
 \int_0^\infty dw\;w^d\ln\brkt{1-e^{-2\pi\sqrt{w^2+\bar{M}^2}}}. 
 \label{E_cas^conv:2}
\eea
We have used the reflection formula for the gamma function in the second line. 

\begin{figure}[t]
\begin{center}
\includegraphics[width=100mm]{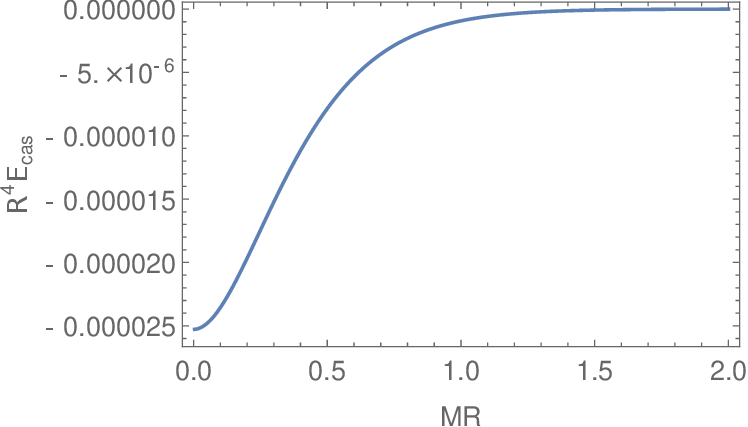}
\end{center}
\caption{$R^4E_{\rm cas}$ in the case of $D=5$ 
as a function of $MR$. }
\label{E_cas^conv}
\end{figure}

The bulk mass~$M$ exponentially suppresses the Casimir Energy as shown in Fig.~\ref{E_cas^conv}. 

\section{Cutoff regularization} \label{cutoff_reg}
Since the expression~(\ref{V_eff:1}) diverges, 
we need some physical condition that extracts a finite physically-sensible contribution, 
\ie, the Casimir energy (density)~$E_{\rm cas}$, 
from the divergent quantity~$E_{\rm vac}$. 
To see this extraction explicitly, 
the dimensional regularization and the zeta function regularization are not suitable. 
Thus we work in the cutoff regularization. 
We introduce the momentum cutoff~$\Lmd$ and the cutoff for the KK masses~$\tl{\Lmd}$, 
independently. 
In contrast to the conventional calculations, we keep them finite. 

As a condition to extract the finite energy density, 
we define the Casimir energy~$E_{\rm cas}$ in such a way that 
it vanishes when the size of the compactification radius is taken to infinite.   
Namely, the energy density (in the $D$-dimensional spacetime) is measured 
from that in the non-compact limit~\cite{Kay:1978zr}. 
In this work, we discuss the global Casimir energy, for simplicity. 
Thus, this condition is adopted to the averaged energy density over the compact space~$E_{\rm vac}/\pi R$. 
Therefore, the Casimir energy~$E_{\rm cas}$, which is defined as the vacuum energy density 
in the $(D-1)$-dimensional effective theory, is defined as 
\be
 \frac{E_{\rm cas}(R)}{\pi R} \equiv \frac{E_{\rm vac}(R)}{\pi R}
 -\lim_{R\to\infty}\frac{E_{\rm vac}(R)}{\pi R}. 
 \label{def:E_cas}
\ee

\subsection{Momentum cutoff} \label{mom_cutoff}
First, we perform the $\vec{k}$-integral. 
\bea
 \int\frac{d^dk}{2(2\pi)^d}\;\brkt{k^2+m_n^2}^{1/2}
 \eql \frac{2\pi^{d/2}}{\Gm(\frac{d}{2})}\int_0^\Lmd\frac{dk}{2(2\pi)^d}\;k^{d-1}\brkt{k^2+m_n^2}^{1/2} \nonumber\\
 \eql \frac{m_n^{d+1}}{2(4\pi)^{d/2}\Gm(\frac{d}{2})}
 \int_{\ep_n}^1 ds\;s^{-\frac{d+1}{2}-1}\brkt{1-s}^{\frac{d}{2}-1} \nonumber\\
 \eql \frac{m_n^{d+1}}{2(4\pi)^{d/2}\Gm(\frac{d}{2})}\bt_{\ep_n}\brkt{-\frac{d+1}{2},\frac{d}{2}}, 
\eea
where $\Lmd$ is the $d$-dimensional momentum cutoff scale, 
$2\pi^{d/2}/\Gm(\frac{d}{2})$ is the area of the $(d-1)$-dimensional sphere with a unit radius, 
\be
 s \equiv \frac{m_n^2}{k^2+m_n^2}, \;\;\;\;\;
 \ep_n \equiv \frac{m_n^2}{\Lmd^2+m_n^2}, 
\ee
and $\bt_z(\alp,\bt)$ is the incomplete beta function defined in (\ref{def:bt:incomplete}). 
Thus, (\ref{V_eff:1}) is expressed as 
\bea
 E_{\rm vac} \eql \frac{1}{2(4\pi)^{d/2}\Gm(\frac{d}{2})}
 \lim_{N\to\infty}\sum_{n=1}^N
 m_n^{d+1}\bt_{\ep_n}\brkt{-\frac{d+1}{2},\frac{d}{2}}. \label{V_eff:2}
\eea
The explicit forms of $\beta_{\ep_n}$ in (\ref{V_eff:2}) for various values of $d$ are collected 
in Appendix~\ref{explicit_bt_ep}. 

If we take the limit~$\Lmd\to\infty$ {\it before} taking the limit~$N\to\infty$, 
the above expression reduces to (\ref{V_eff:DR}) in the previous section 
since~\footnote{
Notice that $B\brkt{-\frac{d+1}{2},\frac{d}{2}}$ is well-defined 
only when $\frac{d+1}{2}$ is not a positive integer. 
In contrast, $\bt_{\ep_n}\brkt{-\frac{d+1}{2},\frac{d}{2}}$ can be defined 
even in such a case as long as $\ep_n$ is kept non-zero. 
} 
\bea
 \lim_{\ep_n\to 0}\bt_{\ep_n}\brkt{-\frac{d+1}{2},\frac{d}{2}}
 = B\brkt{-\frac{d+1}{2},\frac{d}{2}}
 = \frac{\Gm(-\frac{d+1}{2})\Gm(\frac{d}{2})}{\Gm(-\frac{1}{2})}. 
\eea
This indicates that, in the derivation of the Casimir energy shown in the previous section,  
we have implicitly assumed that the contributions from the KK modes near the cutoff~$m_n\sim\Lmd$ 
are negligibly small. 
In the next section, we will numerically check whether this is true or not.

\subsection{Cutoff for KK masses} \label{cutoff_mKK}
Here we introduce the cutoff scale~$\tl{\Lmd}$ for the KK masses. 
Naively, the regularized vacuum energy is written as
\be
 E_{\rm vac} = \frac{1}{2(4\pi)^{d/2}\Gm(\frac{d}{2})}
 \sum_{n=1}^N m_n^{d+1}\bt_{\ep_n}\brkt{-\frac{d+1}{2},\frac{d}{2}}, 
 \label{V_eff:3}
\ee
where 
\be
 N \equiv {\rm floor}\,\brkt{R\sqrt{\tl{\Lmd}^2-M^2}}
\ee 
is the cutoff for the number of the KK modes. 
However, it seems natural to assume that the contributions of the KK modes near the cutoff 
are suppressed by the UV physics.\footnote{
In fact, as we will see later, a sharp cutoff for the KK summation like this leads to 
a divergent Casimir energy in the limit of $\Lmd,\tl{\Lmd}\to\infty$. 
} 
So we introduce the damping function that has the property: 
\be
 g_N(x) \simeq \begin{cases} 1 & \mbox{for $x \ll N$} \\
 0 & \mbox{for $x \gg N$} \end{cases}, \label{prop:gN}
\ee
and consider the following quantity 
instead of (\ref{V_eff:3}). 
\bea
 E_{\rm vac} \eql \frac{1}{2(4\pi)^{d/2}\Gm(\frac{d}{2})}\sum_{n=1}^\infty
 m_n^{d+1}\bt_{\ep_n}\brkt{-\frac{d+1}{2},\frac{d}{2}}g_N(n). 
 \label{expr:E_vac:2}
\eea

\begin{figure}[t]
\begin{center}
\includegraphics[width=160mm]{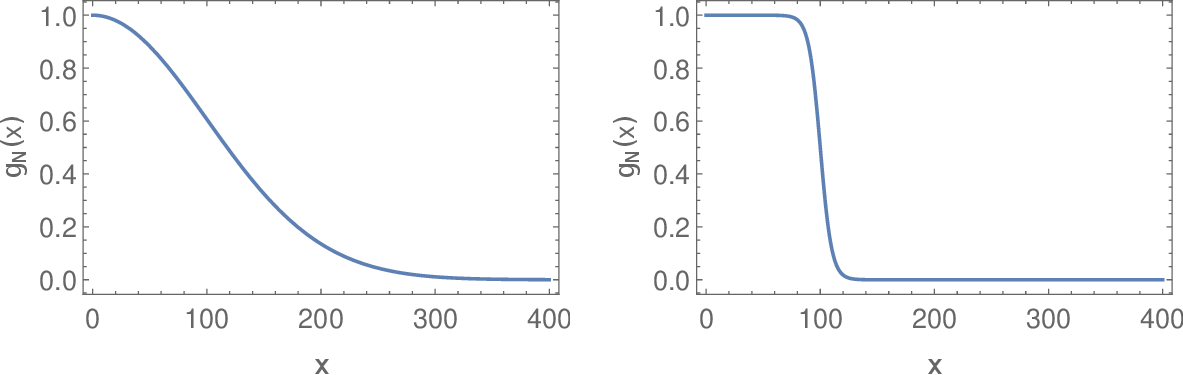}
\end{center}
\caption{The profiles of the damping function~$g_N(x)$ for (\ref{exp:g}) (left plot) 
and for (\ref{kink:g}) (right plot). 
The cutoff~$N$ and the constant~$A$ are chosen as $10^2$ and $10$, respectively. }
\label{gN_prf}
\end{figure}

As examples of the damping function, we can take
\be
 g_N(x) = \exp\brkt{-\frac{x^2}{2N^2}}, \label{exp:g}
\ee
or
\be
 g_N(x) = \frac{1}{2}\sbk{1+\tanh\brc{A\brkt{1-\frac{x}{N}}}}, \label{kink:g}
\ee
where $A$ is a positive constant that controls the steepness around the cutoff (see Fig.~\ref{gN_prf}). 
We cannot choose too small values of $A$. 
Otherwise the damping effect reaches around $x=0$, 
and the property~(\ref{prop:gN}) is no longer satisfied. 
As we will see in the next section, $A\gtrsim 10$ is required.

\begin{figure}[t]
\begin{center}
\includegraphics[width=90mm]{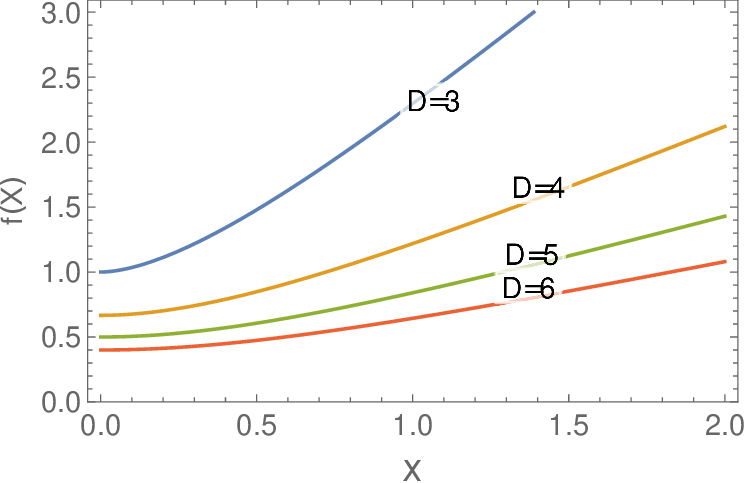}
\end{center}
\caption{The profiles of the function~$f(X)$ in (\ref{expr:fX})
for various dimensions. }
\label{fX_prf}
\end{figure}

We rewrite (\ref{expr:E_vac:2}) in terms of the dimensionless constants:
\be
 a \equiv \frac{1}{\Lmd R}, \;\;\;\;\;
 \hat{M} \equiv \frac{M}{\Lmd},  \label{dimless_pm}
\ee
as
\be
 E_{\rm vac} = \frac{\Lmd^{d+1}}{2(4\pi)^{d/2}\Gm(\frac{d}{2})}\sum_{n=1}^\infty F(an), 
 \label{expr:E_vac:3}
\ee
where
\bea
 F(x) \defa f(X(x))g_N(x), \nonumber\\
 X(x) \defa \sqrt{\hat{M}^2+x^2}, \nonumber\\
 f(X) \defa \bt_{\frac{X^2}{1+X^2}}\brkt{-\frac{d+1}{2},\frac{d}{2}}X^{d+1}. 
 \label{def:F}
\eea
From (\ref{bt_ep:D3})-(\ref{bt_ep:D6}), the explicit forms of the function~$f(X)$ are given by 
\be
 f(X) = \begin{cases} \displaystyle \sqrt{1+X^2}+X^2\ln\frac{1+\sqrt{1+X^2}}{X} 
 & D=3\;\;(d=1) \\
 \displaystyle \rule{0mm}{23pt} \frac{2}{3}\brc{\brkt{1+X^2}^{\frac{3}{2}}-X^3} 
 & D=4\;\;(d=2) \\
 \displaystyle \rule{0mm}{27pt} \frac{1}{4}\brc{\brkt{2+X^2}\sqrt{1+X^2}-X^4\ln\frac{1+\sqrt{1+X^2}}{X}} 
 & D=5\;\;(d=3) \\
 \displaystyle \rule{0mm}{23pt} \frac{2}{15}\brc{\brkt{3-2X^2}\brkt{1+X^2}^{\frac{3}{2}}+2X^5} 
 & D=6\;\;(d=4) \end{cases}. \label{expr:fX}
\ee
Although the explicit functional forms of $f(X)$ for different values of $D$ are quite different, 
they have a similar behavior. 
In fact, we find that 
\be
 f(0) = \frac{2}{D-1}, 
\ee
and
\be
 f(X) \sim \frac{2}{D-2}X. \;\;\;\brkt{\mbox{for $X\gg 1$}}
\ee
Fig.~\ref{fX_prf} shows the profile of $f(X)$ in each dimension. 

Due to the damping function~$g_N(x)$, the function~$F(x)$ satisfies the boundary conditions, 
\be
 \lim_{x\to\infty}F(x) = 0, \;\;\;\;\;
 \lim_{x\to\infty}F^{(p)}(x) = 0. \;\;\; (p=1,2,3,\cdots) \label{prop:F}
\ee

\subsection{Non-compact limit}
Next consider the non-compact limit ($R\to\infty$) 
in order to evaluate the Casimir energy given in (\ref{def:E_cas}). 
When we take this limit, we have to treat the contribution of the zero-mode carefully 
because it is projected out by the orbifold boundary conditions. 
Let us first consider the case of the $S^1$ compactification instead of $S^1/Z_2$. 
Then, the sum in (\ref{expr:E_vac:3}) is taken over the whole integers. 
In this case, we can easily take the non-compact limit. 
\bea
 \lim_{R\to\infty}\frac{E^{S^1}_{\rm vac}(R)}{\pi R}
 \eql \frac{\Lmd^{d+2}}{2\pi(4\pi)^{d/2}\Gm(\frac{d}{2})}
 \lim_{a\to 0}\sum_{n=-\infty}^\infty aF(an)  \nonumber\\
 \eql \frac{\Lmd^{d+2}}{2\pi(4\pi)^{d/2}\Gm(\frac{d}{2})}\int_{-\infty}^\infty dx\;F(x). 
 \label{nc_limit:E/R}
\eea
Thus, we have
\bea
 \frac{E_{\rm cas}^{S^1}}{\pi R} 
 \eql \frac{E^{S^1}_{\rm vac}(R)}{\pi R}-\lim_{R\to\infty}\frac{E^{S^1}_{\rm vac}(R)}{\pi R} 
 \nonumber\\
 \eql \frac{\Lmd^{d+2}}{2\pi(4\pi)^{d/2}\Gm(\frac{d}{2})}
 \brc{\sum_{n=-\infty}^\infty aF(an)-\int_{-\infty}^\infty dx\;F(x)} \nonumber\\
 \eql \frac{\Lmd^{d+2}a}{2\pi(4\pi)^{d/2}\Gm(\frac{d}{2})}
 \brc{\sum_{n=-\infty}^\infty F(an)-\int_{-\infty}^\infty dx\;F(ax)}. 
 \label{E_cas^S1}
\eea
We have used that
\be
 \frac{1}{a}\int_{-\infty}^\infty dx\;F(x) = \int_{-\infty}^\infty dx\;F(ax). 
\ee
Since $F(x)$ is an even function, the brace part in (\ref{E_cas^S1}) can be rewritten as
\be
 \sum_{n=-\infty}^\infty F(an)-\int_{-\infty}^\infty dx\;F(ax) 
 = 2\sum_{n=1}^\infty F(an)+F(0)-2\int_0^\infty dx\;F(ax). 
\ee

Now let us come back to the case of $S^1/Z_2$. 
Noting that (\ref{E_cas^S1}) is finite in the limit of $\Lmd\to\infty$, 
we should identify the non-compact limit of the energy density 
as~\footnote{
For a $Z_2$-even field, this becomes
\be
 \lim_{R\to\infty}\frac{E_{\rm vac}(R)}{\pi R} = \frac{\Lmd^{d+2}}{2\pi(4\pi)^{d/2}\Gm(\frac{d}{2})}
 \brc{\int_0^\infty dx\;F(x)+\frac{1}{2}F(0)}. 
\ee
}
\be
 \lim_{R\to\infty}\frac{E_{\rm vac}(R)}{\pi R}
 = \frac{\Lmd^{d+2}}{2\pi(4\pi)^{d/2}\Gm(\frac{d}{2})}
 \brc{\int_0^\infty dx\;F(x)-\frac{1}{2}F(0)}. 
\ee
Thus the Casimir energy in this case is expressed as
\bea
 E_{\rm cas} \eql \frac{\Lmd^{d+1}}{2(4\pi)^{d/2}\Gm(\frac{d}{2})}\Dlt(a) 
 = \frac{\Dlt(a)/a^{d+1}}{2(4\pi)^{d/2}\Gm(\frac{d}{2})R^{d+1}}, 
 \label{expr:E_cas:Dlt}
\eea
where
\be
 \Dlt(a) \equiv \sum_{n=1}^\infty F(an)-\int_0^\infty dx\;F(ax)+\frac{1}{2}F(0). 
 \label{def:Dlta}
\ee

\ignore{
If we expand $\Dlt(a)$ in (\ref{def:Dlta}) with respect to $a$: 
\be
 \Dlt(a) = C_0+C_1a+C_2a^2+\cdots, 
\ee
the coefficients~$C_i$ ($i\leq d$) need to vanish 
in order for $E_{\rm cas}$ not to diverge in the limit of $\Lmd\to\infty$ (\ie, $a\to 0$). 
The coefficient~$C_{d+1}$ provides the Casimir energy in such a limit.  
}

\section{Calculation of Casimir energy} \label{calc_Cas_eng}
In this section, we numerically calculate the Casimir energy in the cutoff regularization scheme, 
and evaluate the deviation from the conventional result. 
As mentioned in Sec.~\ref{cutoff_mKK}, the behaviors of $f(X)$ in (\ref{expr:fX}) 
are similar for different $D$. 
Thus the qualitative features of the Casimir energy do not 
depend on $D$ very much.  
Therefore we focus on the case of $D=5$ ($d=3$) in this section. 

\subsection{Expression of Casimir energy}
In (\ref{dimless_pm}), we have normalized the bulk mass~$M$ by the cutoff scale~$\Lmd$. 
Since $\hat{M}$ is independent of $R$, it can be treated as a constant when we take 
the non-compact limit ($R\to\infty$). 
However, once we obtain the expression~(\ref{expr:E_cas:Dlt}) with (\ref{def:Dlta}), 
it is more convenient to normalize $M$ by the compactification scale~$1/R$ 
rather than the cutoff scale~$\Lmd$. 
Hence we define 
\be
 \bar{M} \equiv RM, 
\ee
and rewrite $\Dlt(a)$ defined in (\ref{def:Dlta}) as 
\be
 \Dlt(a) = \sum_{n=1}^\infty\bar{F}(n)-\int_0^\infty dx\;\bar{F}(x)+\frac{1}{2}\bar{F}(0), 
 \label{def:Dltabar}
\ee
where 
\bea
 \bar{F}(x) \defa f(a\bar{X}(x))g_N(x)  \nonumber\\
 \bar{X}(x) \defa \sqrt{\bar{M}^2+x^2}. \label{def:barF}
\eea

In order to evaluate $\Dlt(a)$ in (\ref{def:Dltabar}), 
the Euler-Maclaurin formula is useful 
(see Appendix~\ref{EM_fml})~\cite{Boyer:1968uf,Mahajan:2006mw,Saghian:2012zy}. 
Using the formula~(\ref{dif:f}), $\Dlt(a)$ is expressed as 
\bea
 \Dlt(a) \eql \lim_{n_{\rm max}\to\infty}\brc{\sum_{n=1}^{n_{\rm max}} \bar{F}(n)
 -\int_0^{n_{\rm max}} dx\;\bar{F}(x)
 +\frac{1}{2}\bar{F}(0)} \nonumber\\
 \eql \lim_{n_{\rm max}\to\infty}\brc{\sum_{n=0}^{n_{\rm max}} \bar{F}(n)
 -\int_0^{n_{\rm max}} dx\;\bar{F}(x)}
 -\frac{1}{2}\bar{F}(0) \nonumber\\
 \eql \lim_{n_{\rm max}\to\infty}\sbk{\frac{1}{2}\brc{\bar{F}(0)+\bar{F}(n_{\rm max})}
 +\sum_{p=1}^{\fl{q/2}}\frac{B_{2p}}{(2p)!}\brc{\bar{F}^{(2p-1)}(n_{\rm max})-\bar{F}^{(2p-1)}(0)} +R_q} 
 \nonumber\\
 &&-\frac{1}{2}\bar{F}(0) \nonumber\\
 \eql -\sum_{p=1}^{\fl{q/2}}\frac{B_{2p}}{(2p)!}\bar{F}^{(2p-1)}(0)+R_q, 
 \label{Dlt:Deven}
\eea
where $B_{2q}$ are the Bernoulli numbers (see (\ref{B_num})). 
We have used the condition~(\ref{prop:F}) at the last equality. 
The remainder term~$R_q$ is defined as
\be
 R_q \equiv (-1)^{q-1}\int_0^\infty dx\;\frac{B_q(x-\fl{x})}{q!}\bar{F}^{(q)}(x), 
\ee
where $B_q(x)$ is the Bernoulli polynomial (see (\ref{B_poly})), 
and the symbol~$\fl{\cdots}$ denotes the floor function. 
The integer~$q$ can be chosen to an arbitrary value greater than 1. 
Here we set it as $q=2$. 
Then, (\ref{Dlt:Deven}) becomes 
\be
 \Dlt(a) = -\frac{1}{12}\bar{F}^{(1)}(0)+R_2, \label{expr:Dlt}
\ee
and the remainder term is expressed as
\bea
 R_2 \eql -\int_0^\infty dx\;\frac{B_2(x-\fl{x})}{2}\bar{F}^{(2)}(x) \nonumber\\
 \eql -\frac{1}{2}\sum_{k=0}^\infty\int_k^{k+1}dx\;B_2(x-k)\bar{F}^{(2)}(x). 
 \label{expr:R_2}
\eea
Due to the property~(\ref{prop:gN}), we have $g_N(0)\simeq 1$, and thus 
the first term in (\ref{expr:Dlt}) vanishes because 
\be
 \bar{F}^{(1)}(0) \simeq \brc{\der_xf(a\bar{X}(x))}_{x=0} 
 = \brc{\frac{ax}{\bar{X}(x)}f^{(1)}(a\bar{X}(x))}_{x=0} = 0. 
\ee

\subsection{Case of Gaussian damping}
\begin{figure}[t]
\begin{center}
\includegraphics[width=100mm]{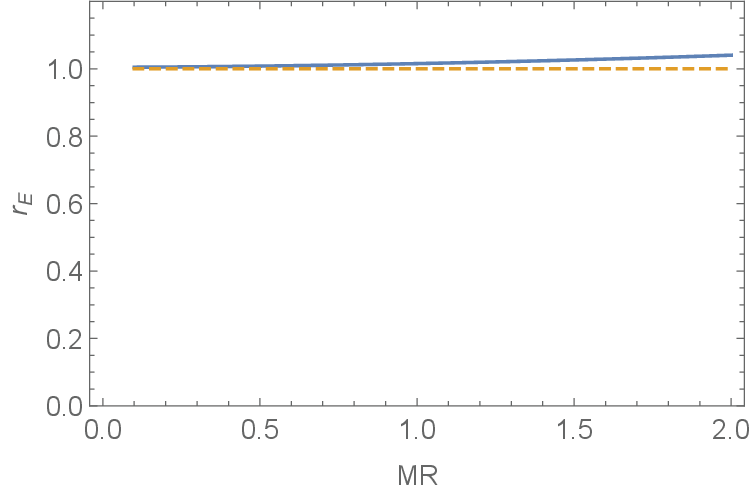}
\end{center}
\caption{The ratio~(\ref{def:r_E}) as a function of the bulk mass~$\bar{M}=MR$. 
The blue solid and orange dashed lines represent the case of 
$\Lmd R=\tl{\Lmd}R=10$ and $100$, respectively. 
}
\label{r_E:exp:g}
\end{figure}

Here we choose the Gaussian-type damping function~(\ref{exp:g}). 
In order to see the deviation from the conventional result, we define the ratio: 
\be
 r_E \equiv \frac{E_{\rm cas}}{E_{\rm cas}^{\rm conv}},  \label{def:r_E}
\ee
where $E_{\rm cas}^{\rm conv}$ is the Casimir energy calculated in the conventional methods, 
such as (\ref{conventional_E_cas}) or (\ref{E_cas^conv:2}). 
Fig.~\ref{r_E:exp:g} shows the ratio~$r_E$ as a function of the bulk mass~$\bar{M}=MR$. 
In this plot, we have chosen the cutoff scales as $\tl{\Lmd}=\Lmd$. 
As this plot shows, the deviation is small and can be neglected if $\Lmd R$ is larger than $10^2$. 
Therefore, the cutoff-dependence of $E_{\rm cas}$ is negligible, 
and the conventional result is reliable even in the case of a finite cutoff scale~$\Lmd$ 
unless $\Lmd R=\cO(10)$. 
This is true also in the case of $\tl{\Lmd}\neq\Lmd$ 
as long as they are in the same order of magnitude.

\subsection{Case of kink-like damping}
\subsubsection{General properties}
\begin{figure}[t]
\begin{center}
\includegraphics[width=100mm]{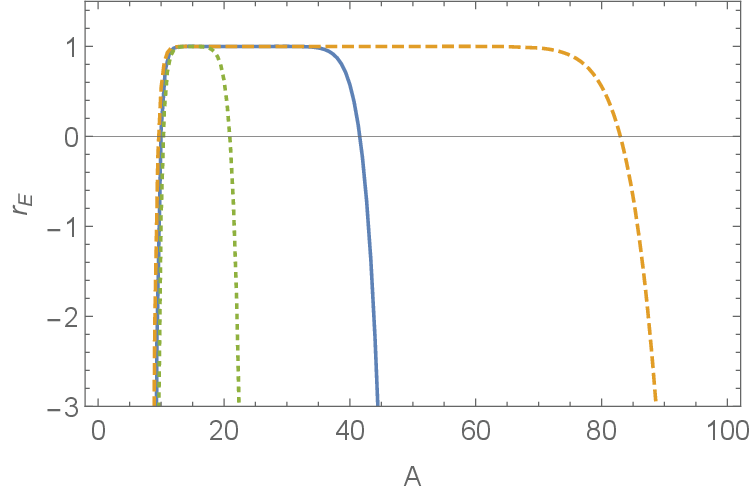}
\end{center}
\caption{The ratio~(\ref{def:r_E}) as a function of $A$ in the damping function~(\ref{kink:g}). 
The cutoff scale is chosen as $a=10^{-2}$, and the bulk mass is chosen as $\bar{M}=0.1$. 
The blue solid, the orange dashed and the green dotted lines represent the case of 
$\tl{\Lmd}=\Lmd$, $2\Lmd$ and $0.5\Lmd$, respectively. }
\label{r_E:kink:g1}
\end{figure}

Next we consider the case of the kink-type damping function~(\ref{kink:g}). 
Fig.~\ref{r_E:kink:g1} shows the ratio~$r_E$ as a function of the parameter~$A$ 
in the damping function. 
The cutoff scale are chosen as $\Lmd R=100$, 
and $\tl{\Lmd}=\Lmd$ (blue solid), $2\Lmd$ (orange dashed) and $0.5\Lmd$ (green dotted). 
The bulk mass is chosen as $\bar{M}=MR=0.1$. 
For $A\lesssim 10$, the ratio~$r_E$ deviates from one because 
the damping effect reaches near the origin and the property~(\ref{prop:gN}) 
is no longer satisfied. 
Namely, $g_N(x)$ is not suitable for the damping function when $A\lesssim 10$. 
In the case of $\tl{\Lmd}=\Lmd$, the result agrees with the conventional one 
for $10\lesssim A\lesssim 35$. 
However, for $A\gtrsim 35$, the ratio~$r_E$ starts to deviate from one, 
flips the sign and becomes negatively larger and larger. 
We can also see that the upper limit of the region in which the result agrees with the conventional one   
is sensitive to the ratio of the cutoffs~$\tl{\Lmd}/\Lmd$.

\subsubsection{The limit of $\bdm{A\to\infty}$} \label{large_A_limit}
\begin{figure}[t]
\begin{center}
\includegraphics[width=75mm]{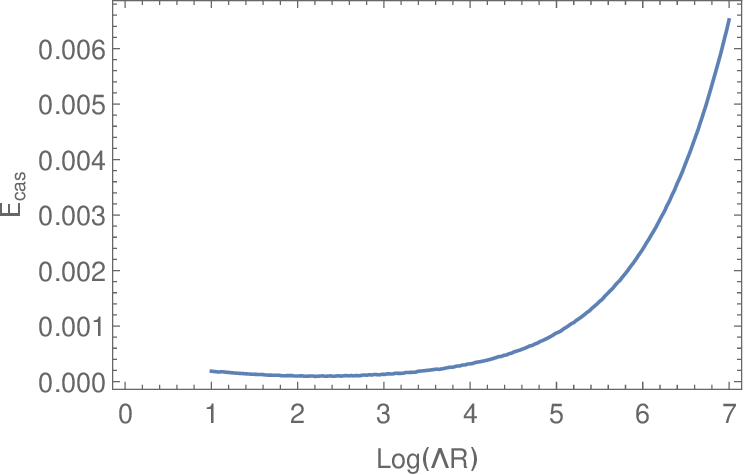} \;\;\;
\includegraphics[width=80mm]{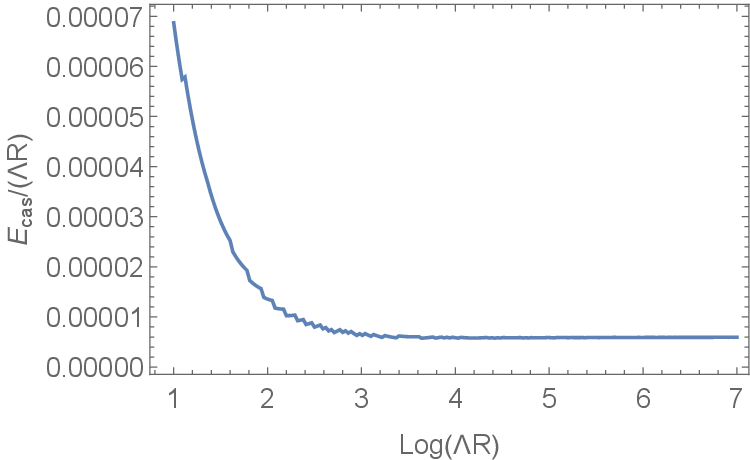}
\end{center}
\caption{The Casimir energy~$E_{\rm cas}$ 
and $E_{\rm cas}/(\Lmd R)$ in the limit of $A\to\infty$ 
as functions of $\ln(\Lmd R)$. }
\label{E_cas-a}
\end{figure}

In order to understand the behavior of $r_E$ for large values of $A$, 
we consider the limit of $A\to\infty$. 
Since 
\bea
 \bar{F}^{(2)}(x) \eql \der_x^2\brc{\cF(x)g_N(x)} \nonumber\\
 \eql \cF^{(2)}(x)g_N(x)+2\cF^{(1)}(x)g_N^{(1)}(x)+\cF(x)g_N^{(2)}(x), 
\eea
where $\cF(x)\equiv f(a\bar{X}(x))$, 
and $g_N^{(1)}(x)$ and $g_N^{(2)}(x)$ sharply localizes around $x=N$ 
when $A$ is large, (\ref{expr:R_2}) is rewritten as 
\bea
 R_2 \eql -\frac{1}{2}\sum_{k=0}^\infty\int_k^{k+1}dx\;B_2(x-k)\cF^{(2)}(x)g_N(x) \nonumber\\
 &&-\frac{1}{2}\sum_{k=N-1}^N\int_k^{k+1}dx\;B_2(x-k)\brc{2\cF^{(1)}(x)g_N^{(1)}(x)
 +\cF(x)g_N^{(2)}(x)}. 
\eea
In the limit of $A\to\infty$, the damping function~$g_N(x)$ and its derivatives become
\bea
 g_N(x) \toa \Tht(N-x), \nonumber\\
 g_N^{(1)}(x) \toa -\dlt(x-N), \nonumber\\
 g_N^{(2)}(x) \toa -\dlt^{(1)}(x-N), 
\eea
where $\Tht(x)$ is the Heaviside step function. 
Therefore, we have
\bea
 &&\sum_{k=N-1}^N\int_k^{k+1}dx\;B_2(x-k)\cF^{(1)}(x)g_N^{(1)}(x) \nonumber\\
 \eql \int_{N-1}^Ndx\;B_2(x-N+1)\cF^{(1)}(x)g_N^{(1)}(x)
 +\int_N^{N+1}dx\;B_2(x-N)\cF^{(1)}(x)g_N^{(1)}(x) \nonumber\\
 \toa -\frac{1}{2}B_2(1)\cF^{(1)}(N)-\frac{1}{2}B_2(0)\cF^{(1)}(N) =- \frac{1}{6}\cF^{(1)}(N), 
\eea
and
\bea
 &&\sum_{k=N-1}^N\int_k^{k+1}dx\;B_2(x-k)\cF(x)g_N^{(2)}(x) \nonumber\\
 \eql \sum_{N-1}^N\brc{\sbk{B_2(x-k)\cF(x)g_N^{(1)}(x)}_k^{k+1}
 -\int_k^{k+1}dx\;\der_x\brc{B_2(x-k)\cF(x)}g_N^{(1)}(x)} \nonumber\\
 \eql B_2\brc{\cF(N+1)g_N^{(1)}(N+1)-\cF(N-1)g_N^{(1)}(N-1)} \nonumber\\
 &&-\sum_{N-1}^N\int_k^{k+1}dx\;\brc{2B_1(x-k)\cF(x)+B_2(x-k)\cF^{(1)}(x)}g_N^{(1)}(x) \nonumber\\
 \toa \frac{1}{2}\brc{2B_1(1)\cF(N)+B_2(1)\cF(N)}
 +\frac{1}{2}\brc{2B_1(0)\cF^{(1)}(N)+B_2(0)\cF^{(1)}(N)} \nonumber\\
 \eql B_2\cF^{(1)}(N) = \frac{1}{6}\cF^{(1)}(N). 
\eea
We have used that 
\bea
 \lim_{A\to\infty}\int_{N-1}^N dx\;\cG(x)g_N^{(1)}(x) \eql -\frac{1}{2}\cG(N), \nonumber\\
 \lim_{A\to\infty}\int_N^{N+1} dx\;\cG(x)g_N^{(1)}(x) \eql -\frac{1}{2}\cG(N), 
\eea
for an arbitrary function~$\cG(x)$, and 
\bea
 \der_xB_2(x) \eql 2B_1(x), \;\;\;\;\;
 \lim_{A\to\infty}g_N^{(1)}(N\pm 1) = 0, \nonumber\\
 B_1(0) \eql -B_1(1) = -\frac{1}{2}, \nonumber\\
 B_2(0) \eql B_2(1)=B_2=\frac{1}{6}. 
\eea
Therefore, since 
\be
 \bar{F}^{(1)}(0) = \cF^{(1)}(0) = \brc{a\bar{X}^{(1)}(x)f^{(1)}(a\bar{X}(x))}_{x=0}
 = \brc{\frac{ax}{\bar{X}(x)}f^{(1)}(a\bar{X}(x))}_{x=0} = 0, 
\ee
(\ref{expr:Dlt}) is expressed as
\bea
 \Dlt(a) \eql R_2 = -\frac{1}{2}\sum_{k=0}^{N-1}\int_k^{k+1}dx\;B_2(x-k)\cF^{(2)}(x)
 +\frac{1}{6}\cF^{(1)}(N)-\frac{1}{2}\brc{\frac{1}{6}\cF^{(1)}(N)} \nonumber\\
 \eql -\frac{1}{2}\sum_{k=0}^{N-1}\int_0^1dx\;B_2(x)\cF^{(2)}(x+k)
 +\frac{1}{12}\cF^{(1)}(N). \label{Dlt:Deven:2}
\eea
In the $D=5$ case, we have
\bea
 \cF^{(1)}(N) \eql a^2N\brc{\sqrt{1+a^2(\bar{M}^2+N^2)}
 -a^2\brkt{\bar{M}^2+N^2}\ln\frac{1+\sqrt{1+a^2(\bar{M}^2+N^2)}}{a\sqrt{\bar{M}^2+N^2}}}, 
 \nonumber\\
 \cF^{(2)}(x) \eql \frac{a^2}{\sqrt{1+a^2(\bar{M}^2+x^2)}}
 \bigg[1 \nonumber\\
 &&\hspace{10mm}
 +a^2\brkt{\bar{M}^2+3x^2}\brc{1-\sqrt{1+a^2(\bar{M}^2+x^2)}
 \ln\frac{1+\sqrt{1+a^2(\bar{M}^2+x^2)}}{a\sqrt{\bar{M}^2+x^2}}}\bigg]. \nonumber\\
\eea
Fig.~\ref{E_cas-a} shows the Casimir energy~$E_{\rm cas}$ 
as a function of $\ln(\Lmd R)$. 
From this plot, we can see that $E_{\rm cas}$ grows as $\cO(10^{-5}\Lmd R)$ 
for $\Lmd R>10^3$. 
Namely, the Casimir energy calculated by using the sharp kink damping function diverges 
as $\Lmd\to\infty$. 
We should also notice that the Casimir energy in this limit~$\lim_{A\to\infty}E_{\rm cas}$ 
is positive, which has the opposite sign to the conventional result.

\subsubsection{Cutoff dependence}
\begin{figure}[t]
\begin{center}
\includegraphics[width=75mm]{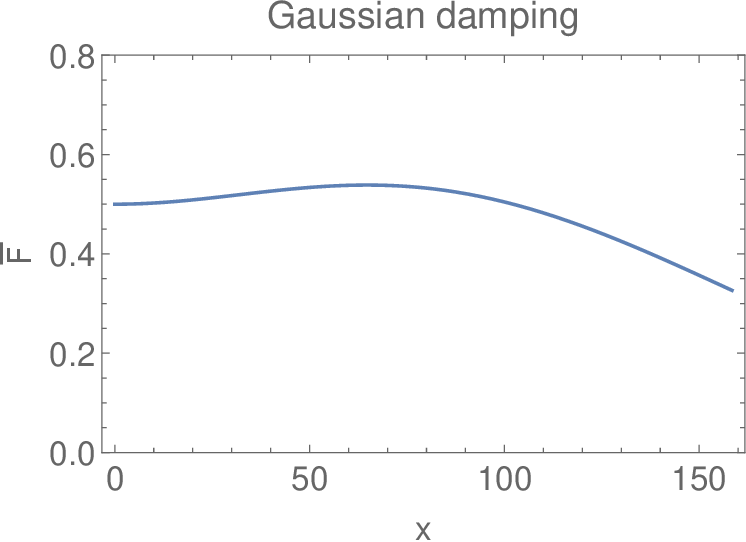} \;\;\;
\includegraphics[width=75mm]{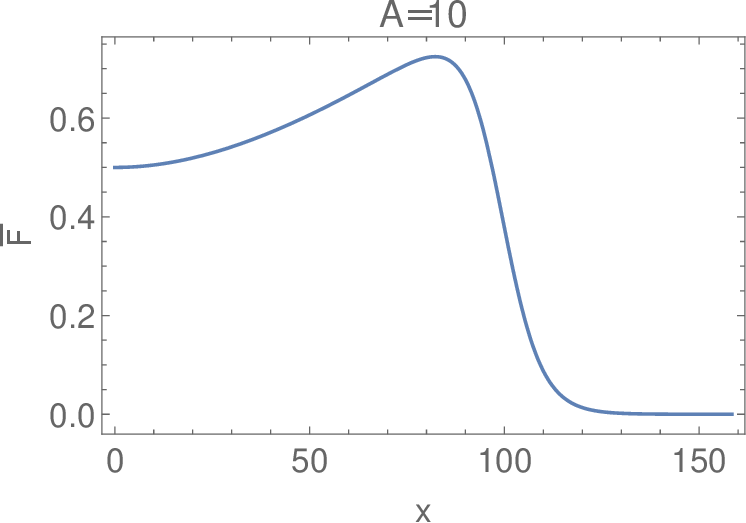}
\end{center}
\begin{center}
\includegraphics[width=75mm]{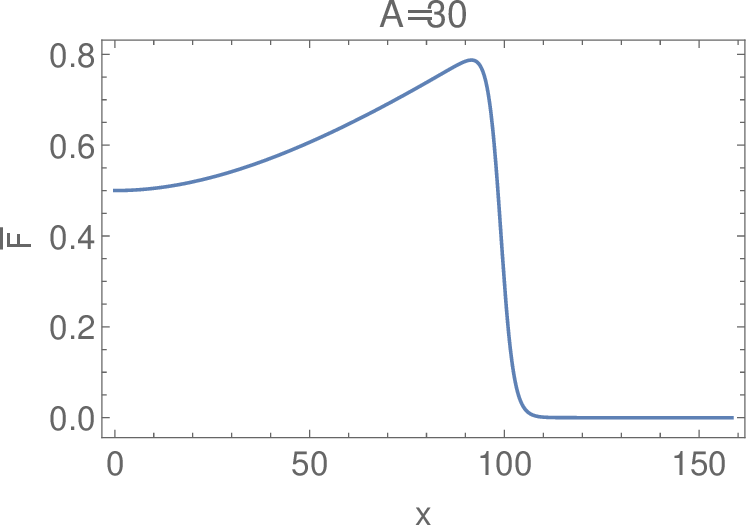} \;\;\;
\includegraphics[width=75mm]{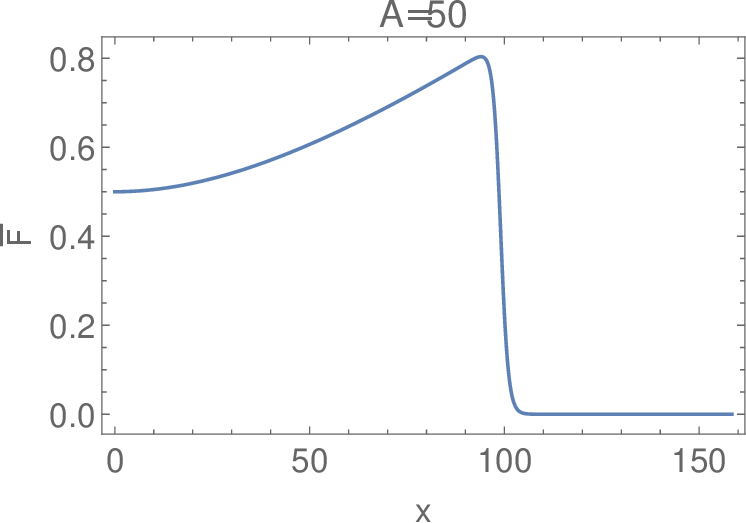}
\end{center}
\caption{The profile of $\bar{F}(x)$ in (\ref{def:barF}) 
in the case of the Gaussian damping function (top left), 
$A=10$ (top right), $A=30$ (bottom left) and $A=50$ (bottom right). 
The bulk mass and the cutoff scale are chosen as $\bar{M}=MR=0.1$ and $\Lmd R=100$, 
respectively. }
\label{Fbar_profiles}
\end{figure}

As we mentioned in Sec.~\ref{mom_cutoff}, the deviation from the conventional result 
comes from the contributions of the KK modes near the cutoff scale~$m_n\sim\Lmd$. 
The Casimir energy~$E_{\rm cas}$ is expressed as
\be
 E_{\rm cas} = \frac{1}{8\pi^2R^4}\frac{\Dlt(a)}{a^4},
\ee
where
\bea
 \Dlt(a) \eql \sum_{n=1}^\infty \bar{F}(n)-\int_0^\infty dx\;\bar{F}(x)+\frac{1}{2}\bar{F}(0). 
\eea
Fig.~\ref{Fbar_profiles} shows the profile of $\bar{F}(x)$ in the case of 
the Gaussian and the kink-like damping functions (the kink-parameter~$A$ is chosen as 10, 30 and 50). 
From these figures, we can see that 
the contributions from the modes near the cutoff scale~$m_n\sim\Lmd$ 
are well suppressed in the case of the Gaussian damping~(\ref{exp:g}). 
Thus the result almost agrees with the conventional one~(\ref{conventional_E_cas}) 
or (\ref{E_cas^conv:2}). 
In contrast, the kink-like damping function~(\ref{kink:g}) takes account of those contributions 
with only small suppression. 
If the kink shape is not very steep, the deviation from the conventional result still remains tiny. 
However, if the damping function becomes close to the step function, 
the deviation becomes large and $E_{\rm cas}$ changes the sign 
as discussed in Sec.~\ref{large_A_limit}. 

\begin{figure}[t]
\begin{center}
\includegraphics[width=75mm]{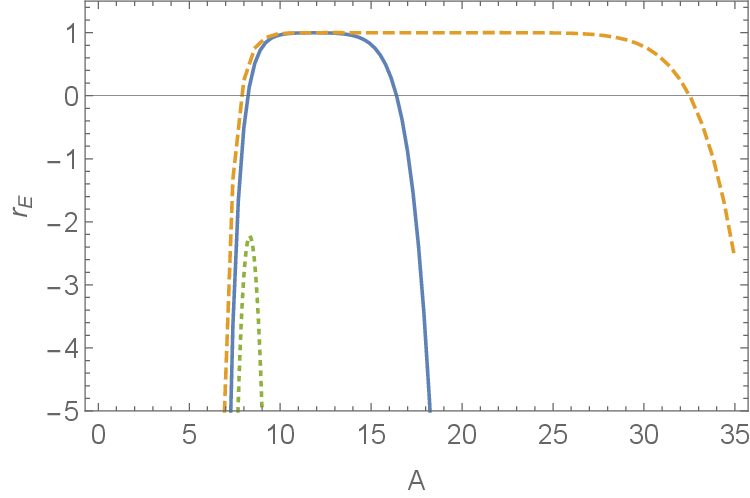} \;\;\;
\includegraphics[width=75mm]{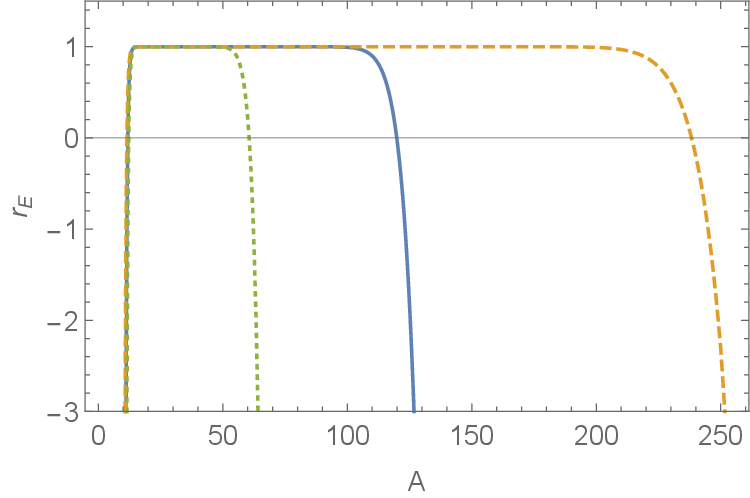}
\end{center}
\caption{The ratio~(\ref{def:r_E}) as a function of $A$ in the damping function~(\ref{kink:g}). 
The cutoff scale is chosen as $a=(\Lmd R)^{-1}=3\times 10^{-2}$ (left plot) 
and $a=3\times 10^{-3}$ (right plot). 
The bulk mass is chosen as $\bar{M}=0.1$. 
The blue solid, the orange dashed and the green dotted lines represent the case of 
$\tilde{\Lmd}=\Lmd$, $2\Lmd$ and $0.5\Lmd$, respectively. }
\label{SLA}
\end{figure}

The range of $A$ in which the conventional result is reproduced 
depends on the hierarchy between the cutoff scale and the compactification scale~$1/R$. 
The plots in Fig.~\ref{SLA} show the ratio~$r_E$ in the cases of $a=3\times 10^{-2}$ (the left plot) 
and of $a=3\times 10^{-3}$ (the right plot). 
We can see that the conventional result is reproduced in a wider range of $A$  
for larger cutoff scales. 
When the cutoff is not high enough, we cannot obtain the conventional result 
for any values of $A$. 

\section{Summary and discussions} \label{summary}
We have numerically estimate the UV cutoff dependence of 
the global Casimir energy in a simple scalar model whose one spatial dimension 
is compactified on $S^1/Z_2$. 
We keep the cutoff scales finite, and evaluate the deviation from the conventional result obtained in the zeta-function regularization, 
which implicitly discards the UV physics information. 
This deviation originates from the contribution of the KK modes near the cutoff scale~$m_n\sim\Lmd$. 
We can easily show that the sharp cutoff for the KK masses leads to a large deviation, 
which changes the sign of the Casimir energy. 
When $\Lmd R>10^3$, this energy is of $\cO(10^{-5}\Lmd R)$. 
Instead of adopting such a sharp cutoff, we introduce a damping function 
and insert it into the infinite KK summation. 
We consider two types of the damping function. 
One is the Gaussian-type and the other is the kink-type. 
In the former case, the contributions near the cutoff scale are well suppressed, 
and the result almost agrees with the conventional one. 
In the latter case, the calculations take into account the modes near the cutoff, 
but the deviation from the conventional result still remains tiny 
unless the kink is extremely steep. 
When the hierarchy between the compactification scale~$1/R$ 
and the cutoff scales~$\Lmd$, $\tl{\Lmd}$ is large, 
the conventional result is reproduced for a wide 
range of the kink parameter~$A$ defined in (\ref{kink:g}). 
However, if the hierarchy is not large enough, such a region becomes narrow 
and the contributions from the KK modes near the cutoff tend to give a non-negligible contribution 
to the Casimir energy. 


Since the 5D extra-dimensional model is non-renormalizable 
and should be regarded as an effective theory, it is expected that 
the contribution of the KK modes to the vacuum energy becomes suppressed 
as the KK mass approaches to the cutoff scale of the 5D theory 
due to the effect of the UV physics. 
Our result indicates that the Casimir energy can receive sizable contribution 
from the modes near the cutoff scale, 
depending on how much the contribution to the vacuum energy is damped near the cutoff. 
In particular, such a deviation from the conventional result tends to appear 
when the cutoff scale is not far from the compactification scale. 

The damping function~$g_N(x)$ introduced in this work mimics 
the suppression of the contributions from higher KK modes by the UV physics. 
For example, let us consider a spontaneously broken supersymmetric (SUSY) model. 
The low-energy effective theory is non-SUSY, and a nonvanishing vacuum energy is induced 
by the non-SUSY field content. 
Above the SUSY-breaking scale~$\Lmd_{\rm SB}$, however, 
the superpartners start to contribute to the vacuum energy, 
and their contributions have opposite signs. 
Therefore, the total contributions from the modes above $\Lmd_{\rm SB}$ are partly cancelled. 
In the non-SUSY effective theory, 
the cancellation by the superpartners is represented 
by the damping function~$g_N(x)$ inserted into the KK summation for the vacuum energy. 
It is interesting to investigate this damping effect in specific models that 
have UV-completed theories. 
For example, we can calculate the Casimir energy in a 6D theory with two extra dimensions 
compactified on a torus. 
If there is a hierarchy between the two radii of the torus, an effective 5D theory appears 
at intermediate scales. 
Then we can explicitly see how the contribution to the vacuum energy from each KK mode behaves 
near the cuoff of the effective 5D theory, and evaluate the deviation from the conventional Casimir 
energy.  

We have worked in the flat spacetime. 
Although the KK mass spectrum depends on the geometry of the extra space, 
the level spacing for the KK masses becomes almost the same  in the UV region, 
\ie, $\Dlt m_n\equiv m_{n+1}-m_n\sim 1/R$. 
Hence the contribution from the KK modes around the cutoff does not depend the geometry 
very much. 
Our result is common in various 5D models. 
However, if the extra space has more dimensions, 
the distributions of the KK mass eigenvalues around the cutoff scale 
are quite different from that of 5D theories. 
Therefore, it is also interesting to extend our work to higher extra dimensions. 

In this work, we have adopted the sharp cutoff for the momentum~$\vec{k}$, for simplicity. 
To investigate the contributions of the modes near the cutoff in more detail, 
we should also introduce the damping function for the momentum integral. 
But we expect that the qualitative features obtained in this work do not change very much.  

We have calculated the one-loop contribution to the Casimir energy in this paper. 
This is enough because we considered the free theory. 
However, once the interaction terms are included, higher-loop contributions 
will appear~\cite{Albrecht:2001cp,DaRold:2003yi,vonGersdorff:2005ce}. 
In such a case, some divergent terms are induced on the orbifold fixed points~\cite{DaRold:2003yi}, 
and thus the corresponding brane terms are need to be introduced in order to renormalize them. 
We can construct a model in which the higher-loop contributions are subdominant
if the cutoff-independence of the Casimir energy is ensured~\cite{vonGersdorff:2005ce}. 
In the case that the cutoff-dependence cannot be neglected, 
it does not obvious whether the cutoff-dependence at two-loop is smaller than that of one-loop. 
We need to check this point when we include the interactions. 

We will discuss these issues in separate papers.

\appendix

\ignore{
\section{Integration formulae}
\bea
 \int\frac{d^dk}{(2\pi)^d}\;\brkt{k^2+m_n^2}^\alp 
 \eql \int\frac{d^dk}{(2\pi)^d}\;\frac{i^{-\alp}}{\Gm(-\alp)}\int_0^\infty dt^{-\alp-1}e^{-i(k^2+m_n^2)t} \nonumber\\
 \eql \frac{i^{-\alp}}{\Gm(-\alp)}\int_0^\infty dt\;t^{-\alp-1}e^{-im_n^2t}\int\frac{d^dk}{(2\pi)^d}e^{-ik^2t} \nonumber\\
 \eql \frac{i^{-\alp}}{\Gm(-\alp)}\int_0^\infty dt\;t^{-\alp-1}e^{-im_n^2t}
 \brkt{\frac{1}{4\pi it}}^{d/2} \nonumber\\
 \eql \frac{i^{-\alp}}{(4\pi i)^{d/2}\Gm(-\alp)}\int_0^\infty dt\;t^{-\alp-1-d/2}e^{-im_n^2t} \nonumber\\
 \eql \frac{i^{-\alp}}{(4\pi i)^{d/2}\Gm(-\alp)}\frac{\Gm(-\alp-d/2)}{(im_n^2)^{-\alp-d/2}} \nonumber\\
 \eql \frac{\Gm(-\alp-d/2)}{(4\pi)^{d/2}\Gm(-\alp)}m_n^{d+2\alp}. \label{int_fml:1}
\eea
We have used that
\bea
 s^\alp \eql \frac{i^{-\alp}}{\Gm(-\alp)}\int_0^\infty dt\;t^{-\alp-1}e^{-ist}, \nonumber\\
 \int d^dk\;e^{-ik^2t} \eql \brkt{\frac{\pi}{it}}^{d/2}. 
\eea
}

\section{Incomplete beta and gamma functions} \label{bega_gamma}
\subsection{Definitions and properties}
The integral expressions of the complete beta and the gamma functions are given by
\bea
 B(\alp,\bt) \defa \int_0^\infty dx\;x^{\alp-1}(1-x)^{\bt-1} = B(\bt,\alp), \nonumber\\
 \Gm(\alp) \defa \int_0^\infty dt\;t^{\alp-1}e^{-t}. 
\eea
These expressions are valid only for $\Re\alp>0$ and $\Re\bt>0$. 
They are related as
\be
 B(\alp,\bt) = \frac{\Gm(\alp)\Gm(\bt)}{\Gm(\alp+\bt)}.  \label{rel:bt-gm}
\ee
This relation holds over the whole domain of the beta function. 

The incomplete beta functions are defined as
\bea
 B_z(\alp,\bt) \defa \int_0^z dx\;x^{\alp-1}(1-x)^{\bt-1}, \nonumber\\
 \bt_z(\alp,\bt) \defa B(\alp,\bt)-B_z(\alp,\bt) = \int_z^1 dx\;x^{\alp-1}\brkt{1-x}^{\bt-1}, 
 \label{def:bt:incomplete}
\eea
and the upper and the lower incomplete gamma functions are defined as
\bea
 \Gm_z(\alp) \defa \int_z^\infty dt\;t^{\alp-1}e^{-t}, \nonumber\\
 \gm_z(\alp) \defa \Gm(\alp)-\Gm_z(\alp) = \int_0^z dt\;t^{\alp-1}e^{-t}. 
\eea
For $\alp+\bt+1>0$ and $\bt>0$, it follows that
\be
 \bt_\ep(\alp,\bt) = \frac{1}{\bt\Gm(\alp+\bt)}\int_0^\infty d\tau\;\tau^\bt e^{-\tau}
 \Gm_{\tau\ep/(1-\ep)}(\alp)+\frac{1}{\bt}\ep^\alp(1-\ep)^\bt. 
\ee
This reduces to (\ref{rel:bt-gm}) in the limit of $\ep\to 0$ 
if $\alp>0$.

\subsection{Explicit forms of Incomplete beta functions} \label{explicit_bt_ep}
The explicit forms of $\beta_{\ep_n}(-\frac{d+1}{2},\frac{d}{2})$, 
where $\ep_n\equiv m_n^2/(\Lmd^2+m_n^2)$, are obtained as follows. 
Since 
\bea
 \der_\Lmd\bt_{\ep_n}\brkt{\alp,\bt}
 \eql -\ep_n^{\alp-1}\brkt{1-\ep_n}^{\bt-1}\der_\Lmd\ep_n \nonumber\\
 \eql \brkt{\frac{m_n^2}{\Lmd^2+m_n^2}}^{\alp-1}\brkt{\frac{\Lmd^2}{\Lmd^2+m_n^2}}^{\bt-1}
 \frac{2\Lmd m_n^2}{\brkt{\Lmd^2+m_n^2}^2} \nonumber\\
 \eql \frac{2\Lmd^{2\bt-1}m_n^{2\alp}}{\brkt{\Lmd^2+m_n^2}^{\alp+\bt}}, 
\eea
and $\bt_{\ep_n}(\alp,\bt)|_{\Lmd=0}=0$, we have
\be
 \bt_{\ep_n}\brkt{-\frac{d+1}{2},\frac{d}{2}}
 = \int_0^\Lmd d\Lmd'\;\der_{\Lmd'}\bt_{\ep_n}\brkt{-\frac{d+1}{2},\frac{d}{2}}
 = \int_0^\Lmd d\Lmd'\;\frac{2\Lmd^{\prime d-1}m_n^{-d-1}}{\brkt{\Lmd^{\prime 2}+m_n^2}^{-\frac{1}{2}}}. 
 \label{int_form:bt_ep}
\ee
Performing the integral, we obtain 
\begin{description}
\item[$\bdm{D=3\;(d=1)}$] 
\bea
 \bt_{\ep_n}\brkt{-1,\frac{1}{2}} \eql \frac{\Lmd\sqrt{\Lmd^2+m_n^2}}{m_n^2}
 +\ln\frac{\Lmd+\sqrt{\Lmd^2+m_n^2}}{m_n}, 
 \label{bt_ep:D3}
\eea

\item[$\bdm{D=4\;(d=2)}$] 
\bea
 \bt_{\ep_n}\brkt{-\frac{3}{2},1} 
 \eql \frac{2\brkt{\Lmd^2+m_n^2}^{\frac{3}{2}}}{3m_n^3}-\frac{2}{3}. 
 \label{bt_ep:D4}
\eea

\item[$\bdm{D=5\;(d=3)}$] 
\bea
 \bt_{\ep_n}\brkt{-2,\frac{3}{2}}
 \eql \frac{\Lmd}{4m_n^4}\brkt{2\Lmd^2+m_n^2}\sqrt{\Lmd^2+m_n^2}
 -\frac{1}{4}\ln\frac{\Lmd+\sqrt{\Lmd^2+m_n^2}}{m_n}. 
 \label{bt_ep:D5}
\eea

\item[$\bdm{D=6\;(d=4)}$] 
\bea
 \bt_{\ep_n}\brkt{-\frac{5}{2},2} 
 \eql \frac{2}{15m_n^5}\brkt{3\Lmd^2-2m_n^2}\brkt{\Lmd^2+m_n^2}^{\frac{3}{2}}+\frac{4}{15}. 
 \label{bt_ep:D6}
\eea
\end{description}

\section{Zeta function regularization} \label{zeta_fct_reg}
\subsection{Integral form}
Here we review the zeta function regularization 
technique~\cite{Garriga:2000jb,Toms:2000bh,Goldberger:2000dv,Brevik:2000vt,Leseduarte:1996xr}. 
Let us consider an infinite sum:
\be
 S_p \equiv \sum_{n=1}^\infty x_n^p, 
\ee
where $x_n$ is a solution of 
\be
 \cF(x_n) = 0. \label{spctrm_eq}
\ee
We assume that all the solutions are located on the positive real axis in the complex plane, and 
\be
 x_n \sim cn, \;\;\;\;\; \brkt{\mbox{$c$: a positive constant}}
\ee
when $n$ is large enough. 
Thus, the infinite sum~$S_p$ converges only when $\Re p<-1$. 
In such a region, $S_p$ can be expressed by a contour integral. 
\bea
 S_p \eql \frac{1}{2\pi i}\oint_C dz\;z^p\frac{\cF'(z)}{\cF(z)} \nonumber\\
 \eql -\frac{p}{2\pi i}\oint_C dz\;z^{p-1}\ln\cF(z), \label{expr:Sp:2}
\eea
where $C$ is a contour that encircles all the solutions of (\ref{spctrm_eq}). 

Suppose that the function~$\cF(x)$ has the following asymptotic behavior. 
\be
 \cF(z) \sim \begin{cases} \cF_+^{\rm asp}(z) & \mbox{for $\Im z>0$} \\
 \cF_-^{\rm asp}(z) & \mbox{for $\Im z<0$} \end{cases}, 
\ee
where $\cF_\pm^{\rm asp}(z)$ are analytic functions that do not have poles 
inside the contour~$C$. 
Then, the expression~(\ref{expr:Sp:2}) can be rewritten as
\bea
 S_p \eql -\frac{p}{2\pi i}\sum_{\sgm=\pm}\int_{C_\sgm}dz\;z^{p-1}\ln\frac{\cF(z)}{\cF_\sgm^{\rm asp}(z)}
 -\frac{p}{2\pi i}\sum_{\sgm=\pm}\int_{C_\sgm}dz\;z^{p-1}\ln\cF^{\rm asp}_\sgm(z), 
 \label{expr:Sp:3}
\eea
where $C_+$ and $C_-$ denote the upper and lower halves of $C$.  
Then, the divergent part is pushed into the second term, which is irrelevant to the physics 
since $\cF^{\rm asp}_\pm(z)$ do not have poles inside $C$.\footnote{
In the case of the Randall-Sundrum background, we need to further extract divergent contribution 
from the first term in (\ref{expr:Sp:3}), which will be renormalized 
to the brane tensions~\cite{Garriga:2000jb}-\cite{Brevik:2000vt}. 
} 

In the case of 
\be
 \cF(z) = \sin\brkt{\pi\sqrt{z^2-\bar{M}^2}}, \label{def:cF}
\ee
where $\bar{M}$ is a positive constant, 
The contour~$C$ consists of 
\bea
 C_1 \eql \brc{z=\cR e^{i\tht}|\tht: 0\to \frac{\pi}{2}} \nonumber\\
 C_2 \eql \brc{z=iw | w: \cR\to \dlt} \nonumber\\
 C_3 \eql \brc{z=x+i\dlt|x: 0\to \bar{M}} \nonumber\\
 C_4 \eql \brc{z=\bar{M}+\dlt e^{i\tht}|\tht: \frac{\pi}{2}\to 0} \nonumber\\
 C_5 \eql \brc{z=\bar{M}+\dlt e^{i\tht}|\tht: 0\to -\frac{\pi}{2}} \nonumber\\
 C_6 \eql \brc{z=x-i\dlt|x: \bar{M}\to 0} \nonumber\\
 C_7 \eql \brc{z=iw|w: -\dlt \to -\cR} \nonumber\\
 C_8 \eql \brc{z=\cR e^{i\tht}|\tht: -\frac{\pi}{2}\to 0}. 
\eea
The asymptotic function~$\cF^{\rm asp}(z)$ is chosen as
\be
 \cF^{\rm asp}(z) = \begin{cases} \displaystyle
 \cF_+^{\rm asp} \equiv \frac{i}{2}\exp\brkt{-i\pi\sqrt{z^2-\bar{M}^2}} & \mbox{for $\Im z>0$} \\
 \displaystyle \rule{0mm}{25pt} \cF_-^{\rm asp} \equiv -\frac{i}{2}\exp\brkt{i\pi\sqrt{z^2-\bar{M}^2}} 
 & \mbox{for $\Im z<0$} 
 \end{cases}. 
\ee
We have chosen the branch of the square root is chosen as 
\be
 \sqrt{re^{i\tht}} = \sqrt{r}e^{i\tht/2}, 
\ee
for $-\pi <\tht\leq \pi$. 
Then, since 
\be
 \frac{\cF(z)}{\cF_\pm^{\rm asp}(z)} = 1-\exp\brkt{\pm 2\pi i\sqrt{z^2-\bar{M}^2}}, 
\ee
we can check that only the contributions from $C_2$ and $C_7$ survive 
in the limit of $\cR\to\infty$ and $\dlt\to 0$. 
Thus, we have
\bea
 S_p \eql -\frac{p}{2\pi i}\brc{\int_{C_2}dz\;z^{p-1}\ln\frac{\cF(z)}{\cF_+^{\rm asp}(z)}
 +\int_{C_7}dz\;z^{p-1}\ln\frac{\cF(z)}{\cF_-^{\rm asp}(z)}}+\cdots \nonumber\\
 \eql -\frac{p}{2\pi i}\left\{\int_\infty^0 d(iw)\;(iw)^{p-1}\ln\brkt{1-e^{-2\pi\sqrt{w^2+\bar{M}^2}}}
 \right.\nonumber\\
 &&\hspace{13mm}\left. 
 +\int_0^{-\infty}d(iw)\;(iw)^{p-1}\ln\brkt{1-e^{-2\pi\sqrt{w^2+\bar{M}^2}}}\right\}+\cdots \nonumber\\
 \eql \frac{p}{\pi}\sin\frac{p\pi}{2}\int_0^\infty dw\;w^{p-1}\ln\brkt{1-e^{-2\pi\sqrt{w^2+\bar{M}^2}}}
 +\cdots, \label{int_form:Sp}
\eea
where the ellipsis denotes divergent terms that are irrelevant to the physics.

\subsection{Infinite-sum form}
In the case of (\ref{def:cF}), there is another expression 
for $S_p$~\cite{Ambjorn:1981xw}-\cite{Nesterenko:1997ku}. 
In this case, $S_p$ becomes
\be
 S_p = \sum_{n=1}^\infty \brkt{n^2+\bar{M}^2}^{p/2}. 
\ee
This converges only when $\Re p<-1$. 

Using the integral representation for the gamma function: 
\be
 \brkt{n^2+\bar{M}^2}^{p/2}\Gm\brkt{-\frac{p}{2}} 
 = \int_0^\infty dt\;t^{-\frac{p}{2}-1}e^{-(n^2+\bar{M}^2)t}, 
\ee
which is valid for $\Re p<0$, it is rewritten as
\bea
 S_p \eql \frac{1}{\Gm(-\frac{p}{2})}\sum_{n=1}^\infty\int_0^\infty dt\;t^{-\frac{p}{2}-1}e^{-(n^2+\bar{M}^2)t} \nonumber\\
 \eql \frac{1}{\Gm(-\frac{p}{2})}\int_0^\infty dt\;t^{-\frac{p}{2}-1}e^{-t\bar{M}^2}\vth(t), 
 \label{New_expr:Sp:2}
\eea
where $\vth(t)\equiv \sum_{n=1}^\infty e^{-n^2t}$ is the Jacobi theta function, and has the property: 
\be
 \vth(t) = -\frac{1}{2}+\frac{1}{2}\sqrt{\frac{\pi}{t}}+\sqrt{\frac{\pi}{t}}\vth\brkt{\frac{\pi^2}{t}}. \label{property:vth}
\ee
Substituting this into (\ref{New_expr:Sp:2}), we obtain
\bea
 S_p \eql -\frac{\bar{M}^p}{2}+\frac{\sqrt{\pi}\Gm(-\frac{p+1}{2})}{2\Gm(-\frac{p}{2})}\bar{M}^{p+1} \nonumber\\
 &&+\frac{\sqrt{\pi}}{\Gm(-\frac{p}{2})}\sum_{n=1}^\infty\int_0^\infty dt\;t^{-\frac{p+3}{2}}
 \exp\brkt{-t\bar{M}^2-\frac{\pi^2n^2}{t}}.  \label{New_expr:Sp:3}
\eea
Note that the $t$-integrals of the contributions from the first and the second terms in (\ref{property:vth}) 
converge only when $\Re p<0$ and $\Re p<-1$, respectively. 
However, once they are expressed with respect to the gamma function, 
they can be analytically connected to the whole complex plane 
except for the points~$p=1,3,5,\cdots$.\footnote{
At the points~$p=0,2,4,\cdots$, the second term in (\ref{New_expr:Sp:3}) simply vanises. 
}
The $t$-integral in the last term of (\ref{New_expr:Sp:3}) converges for any values of $p$, 
thanks to the exponential factor~$\exp\brkt{-t\bar{M}^2-\frac{\pi^2n^2}{t}}$. 
In fact, it is expressed with respect to the modified Bessel function by using the integral expression: 
\be
 K_\alp(z) = \frac{1}{2}\brkt{\frac{z}{2}}^\alp\int_0^\infty dt\;t^{-\alp-1}
 \exp\brkt{-t-\frac{z^2}{4t}}, 
\ee
which is valid for $\Re\alp > -1/2$ and $\abs{\arg z}<\pi/4$. 
Thus, we can analytically continue the above expression to the expression, 
\be
 S_p = -\frac{\bar{M}^p}{2}+\frac{\sqrt{\pi}\Gm(-\frac{p+1}{2})}{2\Gm(-\frac{p}{2})}\bar{M}^{p+1} 
 +\frac{2\bar{M}^{\frac{p+1}{2}}}{\pi^{\frac{p}{2}}\Gm(-\frac{p}{2})}\sum_{n=1}^\infty
 n^{-\frac{p+1}{2}}K_{\frac{p+1}{2}}\brkt{2\pi n\bar{M}}, \label{final_expr:Sp:2}
\ee
which is valid for any values of $p$ except for $p=1,3,5,\cdots$. 
In this expression, the infinite sum in the last term rapidly converges.

\section{Subtraction Scheme} \label{subtract_fml}
\subsection{Euler-Maclaurin formula} \label{EM_fml}
For arbitrary integers~$n_{\rm min}$ and $n_{\rm max}$ ($n_{\rm min}<n_{\rm max}$), 
a $q$ times continuously differentiable function~$h(x)$ on the interval~$[n_{\rm min},n_{\rm max}]$ satisfies
\bea
 &&\sum_{n=n_{\rm min}}^{n_{\rm max}}h(n)
 -\int_{n_{\rm min}}^{n_{\rm max}}dx\;h(x) \nonumber\\
 \eql \frac{1}{2}\brc{h(n_{\rm min})+h(n_{\rm max})}
 -\sum_{k=2}^q (-1)^{k-1}\frac{B_k}{k!}\brc{h^{(k-1)}(n_{\rm max})-h^{(k-1)}(n_{\rm min})}+R_q \nonumber\\
 \eql \frac{1}{2}\brc{h(n_{\rm min})+h(n_{\rm max})}
 +\sum_{p=1}^{\fl{q/2}}\frac{B_{2p}}{(2p)!}\brc{h^{(2p-1)}(n_{\rm max})
 -h^{(2p-1)}(n_{\rm min})}+R_q, \label{dif:f}
\eea
where $B_k$ ($k=2,3,\cdots$) are the Bernoulli numbers, 
\bea
 B_2 \eql \frac{1}{6}, \;\;\;\;\;
 B_4 = -\frac{1}{30}, \;\;\;\;\;
 B_6 = \frac{1}{42}, \;\;\;\;\;
 \cdots \nonumber\\
 B_{2p+1} \eql 0, \;\;\;\;\; \brkt{p=1,2,3,\cdots} 
 \label{B_num}
\eea
and 
\be
 R_q \equiv (-1)^{q-1}\int_{n_{\rm min}}^{n_{\rm max}}dx\;\frac{B_q(x-\fl{x})}{q!}h^{(q)}(x). 
\ee
The symbole~$\fl{\cdots}$ denotes the floor function. 
Here, $B_q(x)$ is the Bernoulli polynomial whose explicit expression is given by 
\bea
 B_0(x) \eql 1, \nonumber\\
 B_1(x) \eql x-\frac{1}{2}, \nonumber\\
 B_2(x) \eql x^2-x+\frac{1}{6}, \nonumber\\
 B_3(x) \eql x^3-\frac{3}{2}x^2+\frac{1}{2}x, \nonumber\\
 B_4(x) \eql x^4-2x^3+x^2-\frac{1}{30}, \nonumber\\
 B_5(x) \eql x^5-\frac{5}{2}x^4+\frac{5}{3}x^3-\frac{1}{6}x, \nonumber\\
 B_6(x) \eql x^6-3x^5+\frac{5}{2}x^4-\frac{1}{2}x^2+\frac{1}{42}, \nonumber\\
 &\vdots&  \label{B_poly}
\eea
These polynomials satisfy the relation
\be
 B_q(1-x) = (-1)^qB_q(x). \label{B:reflection}
\ee
\ignore{
Since 
\be
 \abs{B_q(x)} \leq \frac{2q!}{(2\pi)^q}\zeta(q), 
\ee
for $q\geq 0$ and $0\leq x\leq 1$, 
the remainder term~$R_q$ is bounded as
\be
 \abs{R_q} \leq \frac{2\zeta(q)}{(2\pi)^q}\int_{n_{\rm min}}^{n_{\rm max}} dx\;\abs{h^{(q)}(x)}. 
 \label{rem_term:BD}
\ee
}

\subsection{Simple examples} \label{simple_eg}
Let us define   
\be
 H[f](N) \equiv \sum_{n=0}^N f(n)-\int_0^Ndx\; f(x), \label{def:Hh}
\ee
where 
\be
 f(x) = x, \;\; x^2, \;\; x^3. 
\ee
Then we can explicitly calculate $H[f](N)$ as
\bea
 H[x](N) \eql \sum_{n=0}^N n -\int_0^N dx\;x 
 = \frac{N(N+1)}{2}-\frac{N^2}{2} = \frac{N}{2}, \nonumber\\
 H[x^2](N) \eql \sum_{n=0}^N n^2 -\int_0^N dx\;x^2
 = \frac{N(N+1)(2N+1)}{6}-\frac{N^3}{3} = \frac{3N^2+N}{6}, \nonumber\\
 H[x^3](N) \eql \sum_{n=0}^N n^3-\int_0^N dx\;x^3
 = \frac{N^2(N+1)^2}{4}-\frac{N^4}{4} = \frac{2N^3+N^2}{4}. 
 \label{explicit_eg}
\eea
These results can be reproduced by means of the Euler-Maclaurin formula. 
From (\ref{dif:f}), we have 
\be
 H[f](N) = \frac{1}{2}\brc{f(0)+f(N)}+\sum_{p=1}^{\fl{(m+1)/2}}\frac{B_{2p}}{(2p)!}
 \brc{f^{(2p-1)}(N)-f^{(2p-1)}(0)}+R_{m+1}, \label{dif:f:eg}
\ee
where we have chosen the integer~$q$ in (\ref{dif:f}) as $q=m+1$. 
Since $f^{(m+1)}(x)=0$, the remainder term~$R_{m+1}$ vanishes. 
Therefore, we have 
\bea
 H[x](N) \eql \frac{1}{2}\brc{f(0)+f(N)}+\frac{B_2}{2}\brc{f^{(1)}(N)-f^{(1)}(0)} \nonumber\\
 \eql \frac{1}{2}\brkt{0+N}+\frac{1}{12}\brkt{1-1} = \frac{N}{2}, \nonumber\\
 H[x^2](N) \eql \frac{1}{2}\brc{f(0)+f(N)}+\frac{B_2}{2}\brc{f^{(1)}(N)-f^{(1)}(0)} \nonumber\\
 \eql \frac{1}{2}\brkt{0+N^2}+\frac{1}{12}\brkt{2N-0} = \frac{N^2}{2}+\frac{N}{6}, \nonumber\\
 H[x^3](N) \eql \frac{1}{2}\brc{f(0)+f(N)}+\frac{B_2}{2}\brc{f^{(1)}(N)-f^{(1)}(0)} 
 +\frac{B_4}{4!}\brc{f^{(3)}(N)-f^{(3)}(0)} \nonumber\\
 \eql \frac{1}{2}\brkt{0+N^3}+\frac{1}{12}\brkt{3N^2-0}-\frac{1}{720}\brkt{6-6} 
 = \frac{N^3}{2}+\frac{N^2}{4}. 
 \label{explicit_eg2}
\eea
These agrees with (\ref{explicit_eg}).

\ignore{
\subsection{Introduction of damping function}
In order to remove the remaining would-be divergent terms in (\ref{explicit_eg}) (or (\ref{explicit_eg2})), 
we introduce the damping function: 
\be
 g_N(x) \equiv \frac{1}{2}\sbk{1+\tanh\brc{A\brkt{1-\frac{x}{N}}}}, 
\ee
where $A$ is a positive number greater than one, e.g., $A=10$, and define 
\be
 h(x;N) \equiv f(x)g_N(x). 
\ee
If $A$ is large enough, this behaves as
\be
 h(x;N) \simeq \begin{cases} f(x) & x<N \\
 0 & x >N \end{cases}, \label{app_hxN}
\ee
because 
\be
 \lim_{A\to\infty}g_N(x) = \Tht(N-x), 
\ee
where $\Tht$ is the Heaviside step function. 
With the help of the damping function, we modify (\ref{def:Hh}) as 
\be
 \tl{H}[f](N) \equiv \sum_{n=0}^\infty h(n;N)-\int_0^\infty dx\;h(x;N).  \label{def:tlHf}
\ee
For a large value of $A$, this is approximately 
\bea
 \tl{H}[f](N) \sma \sum_{n=0}^{N-1}f(n)+\frac{1}{2}f(N)-\int_0^N dx\;f(x),
\eea
since $g_N(N)=1/2$. 
Namely, we have
\bea
 \tl{H}[x](N) \sma \frac{N(N-1)}{2}+\frac{N}{2}-\frac{N^2}{2} = 0, \nonumber\\
 \tl{H}[x^2](N) \sma \frac{N(N-1)(2N-1)}{6}+\frac{N^2}{2}-\frac{N^3}{3} 
 = \frac{N}{6}, \nonumber\\
 \tl{H}[x^3](N) \sma \frac{N^2(N-1)^2}{4}+\frac{N^3}{2}-\frac{N^4}{4}
 = \frac{N^2}{4}. \label{subsubleading}
\eea
Now the subleading terms that survived in (\ref{explicit_eg}) are cancelled in this modified quantity, 
but there still remain some divergent terms in the limit of $N\to\infty$. 
Next we will see that from which term in the Euler-Maclaurin formula do these would-be divergent 
terms come. 
Due to the damping function~$g_N(x)$, note that
\be
 \lim_{x\to\infty}h(x;N) = 0. 
\ee
Hence the formula becomes 
\bea
 \tl{H}[f](N) \eql \frac{1}{2}h(0;N)-\sum_{p=1}^{\fl{(m+1)/2}}\frac{B_{2p}}{(2p)!}h^{(2p-1)}(0;N)+R_{m+1}. 
 \label{eg:Hh}
\eea
If we choose the constant~$A$ large enough, 
\bea
 \tl{H}[x](N) \eql \frac{1}{2}f(0)-\frac{B_2}{2}f^{(1)}(0)+R_2 = -\frac{1}{12}+R_2, \nonumber\\
 \tl{H}[x^2](N) \eql \frac{1}{2}f(0)-\frac{B_2}{2}f^{(1)}(0)+R_3 = R_3, \nonumber\\
 \tl{H}[x^3](N) \eql \frac{1}{2}f(0)-\frac{B_2}{2}f^{(1)}(0)-\frac{B_4}{4!}f^{(3)}(0)+R_4 
 = \frac{1}{120}+R_4. 
\eea
By comparing these with (\ref{subsubleading}), 
we find that
\bea
 R_2 \eql \frac{1}{12}, \;\;\;\;\;
 R_3 = \frac{N}{6}, \;\;\;\;\;
 R_4 = -\frac{1}{120}+\frac{N^2}{4}. 
\eea
Thus, for $m\geq 2$, the remainder term diverges as $N\to\infty$. 
If we choose $A$ large enough to 
}

\ignore{
By comparing these with (\ref{subsubleading}), we can see that the would-be divergent terms 
are contained in the remainder terms~$R_{m+1}$. 
In contrast to the case of $H[f](N)$, the remainder term is now nonvanishing 
due to the presence of the damping function~$g_N(x)$. 
\bea
 R_{m+1} \eql (-1)^m\int_0^\infty dx\;\frac{B_{m+1}(x-\fl{x})}{(m+1)!}h^{(m+1)}(x;N). 
 \label{expr:R_m1}
\eea
Notice that the function: 
\be
 h^{(m+1)}(x) = \der_x^{m+1}\brc{x^m g_N(x)} = (m+1)!g_N^{(1)}(x)+\cdots +x^m g_N^{(m+1)}, 
 \label{h^m1}
\ee
takes non-negligible values only around $x=N$. 
Thus this picks up the information near the cutoff~$N$. 
Therefore, if we modify the definition of $\tl{H}[f](N)$ in (\ref{def:tlHf}) as 
\bea
 \hat{H}[x^m](N) \defa \tl{H}[x^m](N)-R_{m+1}, 
\eea
we have finite results in the limit of $N\to\infty$. 
\bea
 \hat{H}_2[x](N) \eql -\frac{1}{12} = \zeta(-1), \nonumber\\
 \hat{H}_3[x^2](N) \eql 0 = \zeta(-2), \nonumber\\
 \hat{H}_4[x^3](N) \eql \frac{1}{120} = \zeta(-3). 
\eea
Namely, this operation is equivalent to the zeta-function regularization 
in the limit of $N\to\infty$. 
}

\ignore{
\subsection{General case}
For a general increasing function~$f(x)$, (\ref{h^m1}) becomes 
\bea
 h^{(q)}(x;N) \eql f^{(q)}(x)g_N(x)+qf^{(q-1)}(x)g_N^{(1)}(x)+\cdots+f(x)g_N^{(q)}(x) \nonumber\\
 \defa f^{(q)}(x)g_N(x)+\Xi_q[f](x). \label{der_h}
\eea
We choose $q$ in such a way that $f^{(q)}(x)$ is the first term in (\ref{der_h}) 
does not provide a non-negligible $N$-dependent contribution to 
the remainder term~$R_q$. 
Then we modify $\tl{H}[f](N)$ in (\ref{def:tlHf}) as 
\be
 \hat{H}_q[f](N) \equiv \sum_{n=0}^\infty f(x)g_N(x)
 -\int_0^\infty dx\;\brc{f(x)g_N(x)+(-1)^{q-1}\frac{B_q(x-\fl{x})}{q!}\Xi_q[f](x)}, 
\ee
so that this remains finite in the limit of $N\to\infty$. 
Then, since $g_N(x)\to\Tht(N-x)$ as $A\to\infty$, 
the Euler-Maclaurin formula indicates that 
\be
 \hat{H}_q[f](N) = \frac{1}{2}f(0)-\sum_{p=1}^{\fl{q/2}}\frac{B_{2p}}{(2p)!}f^{(2p-1)}(0)+\hat{R}_q(N), 
 \label{expr:hatH}
\ee
where
\be
 \hat{R}_q(N) \equiv \frac{(-1)^{q-1}}{q!}\int_0^N dx\;B_q(x-\fl{x})f^{(q)}(x). \label{hatRterm}
\ee
Therefore, (\ref{expr:R_m1}) can be well approximated as
\bea
 R_{m+1} \sma \frac{(-1)^m}{(m+1)!}\int_0^1 dx\;
 B_{m+1}(x)\brc{h^{(m+1)}(x+N-1)+h^{(m+1)}(x+N)} \nonumber\\
 \eql \frac{(-1)^m}{(m+1)!}\left\{\int_0^1 d\tl{x}\;B_{m+1}(1-\tl{x})h^{(m+1)}(N-\tl{x})
 +\int_0^1dx\;B_{m+1}(x)h^{(m+1)}(N+x)
 \right\} \nonumber\\
 \eql \frac{(-1)^m}{(m+1)!}\int_0^1dx\;B_{m+1}(x)\brc{(-1)^{m+1}h^{(m+1)}(N-x)
 +h^{(m+1)}(N+x)}, 
\eea
where $\tl{x}\equiv 1-x$. 
We have used (\ref{B:reflection}) at the last equality. 
Since 
\bea
 h^{(2)}(x) \eql \der_x^2\brc{xg_N(x)} 
 = -\frac{A}{\cosh^2\brc{A\brkt{N-x}}}\sbk{1+Ax\tanh\brc{A\brkt{N-x}}}, \nonumber\\
 h^{(3)}(x) \eql \der_x^3\brc{x^2g_N(x)} \nonumber\\
 \eql -\frac{A\brkt{3+2A^2x^2}}{\cosh^2\brc{A\brkt{N-x}}}
 +\frac{3A^3x^2}{\cosh^4\brc{A\brkt{N-x}}}
 -\frac{6A^2x\tanh\brc{A\brkt{N-x}}}{\cosh^2\brc{A\brkt{N-x}}}, 
 \nonumber\\
 h^{(4)}(x) \eql \der_x^4\brc{x^3g_N(x)} \nonumber\\
 \eql -\frac{9A\brkt{1-2A^2x^2}}{\cosh^4\brc{A\brkt{N-x}}}
 -\frac{3A\brkt{1+2A^2x^2}\cosh\brc{3A\brkt{N-x}}}{\cosh^5\brc{A\brkt{N-x}}} \nonumber\\
 &&-\frac{2A^2x\sinh\brc{A\brkt{N-x}}}{\cosh^5\brc{A\brkt{N-x}}}
 \brc{9-5A^2x^2+\brkt{9+A^2x^2}\cosh\brc{2A\brkt{N-x}}}, \nonumber\\
\eea
we can evaluate $R_{m+1}$ as 
\bea
 R_2 \sma -\frac{1}{2}\int_0^1dx\;B_2(x)\brc{h^{(2)}(N-x)+h^{(2)}(N+x)} \nonumber\\
 \eql \int_0^1dx\;B_2(x)\frac{A\brc{1-Ax\tanh(Ax)}}{\cosh^2(Ax)} 
 \to \frac{1}{12}, \nonumber\\
 R_3 \sma \frac{1}{6}\int_0^1 dx\;B_3(x)\brc{-h^{(3)}(N-x)+h^{(3)}(N+x)} \nonumber\\
 \eql N\int_0^1 dx\;B_3(x)\frac{A^2\brc{4Ax-2Ax\cosh(2Ax)+3\sinh(2Ax)}}{3\cosh^4(A x)} 
 \to \frac{N}{6}, \nonumber\\
 R_4 \sma -\frac{1}{24}\int_0^1dx\;B_4(x)\brc{h^{(4)}(N-x)+h^{(4)}(N+x)} \nonumber\\
 \eql \int_0^1 dx\;B_4(x)\left\{
 \frac{3A\brkt{1-2A^2(N^2+x^2)}}{4\cosh^4(Ax)}
 +\frac{A\brkt{1+2A^2(N^2+x^2)}\cosh(3Ax)}{4\cosh^5(Ax)} \right.\nonumber\\
 &&\hspace{20mm}\left.
 -\frac{A^2x\brc{9-5A^2(3N^2+x^2)+\brkt{9+A^2(3N^2+x^2)}\cosh(2Ax)}\sinh(Ax)}{6\cosh^5(Ax)}
 \right\} \nonumber\\
 \toa \frac{N^2}{4}+0.00342\cdots. 
\eea
We have taken the limit~$A\to \infty$ at the last steps. 
}


\end{document}